\documentclass[aps,twocolumn,longbibliography,nofootinbib,showkeys,showpacs]{revtex4-2}
\usepackage[latin1,utf8]{inputenc} 
\usepackage{amssymb}
\usepackage{amsfonts} 
\usepackage{graphicx} 
\usepackage{float}
\usepackage{amsthm} 
\usepackage{amsmath} 
\usepackage{bm}
\usepackage{epsfig}
\usepackage{hyperref}
\usepackage{color}
\usepackage{bbold}
 \usepackage{bbm}
\usepackage[T1]{fontenc}
\usepackage{young}
\usepackage{youngtab}
\usepackage{xcolor}
\usepackage{braket}

\def\nn{\nonumber}
\def\ic{\mathrm{i}}

\def \bc {\begin{center}}
\def \ec {\end{center}}
\def \bi {\begin{itemize}}
\def \ei {\end{itemize}}

\def \ba {\begin{array}}
\def \ea {\end{array}}

\def \be {\begin{equation}}
\def \ee {\end{equation}}

\newcommand{\la}{\langle}
\newcommand{\ra}{\rangle}

\def\tr {\mathrm{tr}}

\def\bmm {\bm{m}}

\begin{document}

\title{Capturing magic angles in twisted bilayer graphene from information theory markers}

\author{Manuel Calixto}
\email{calixto@ugr.es}
\affiliation{Departamento de Matemática Aplicada, Universidad de Granada, 
	Fuentenueva s/n, 18071 Granada, Spain}
\affiliation{Instituto Carlos I de Física Teórica y  Computacional (iC1), Universidad de Granada, 
	Fuentenueva s/n, 18071 Granada, Spain}

\author{Alberto Mayorgas}
\email{albmayrey97@gmail.com}
\affiliation{Departamento de Ingeniería, Universidad CEU Fernando III,  
Glorieta Cardenal Herrera Oria, 41930 Bormujos, Sevilla, 
Spain}

\author{Octavio Casta\~nos}
\email{ocasta@nucleares.unam.mx}
\affiliation{Instituto de Ciencias Nucleares, Universidad Nacional Autónoma de  Mexico, Apdo.\ Postal 70-543, 04510, CDMX, Mexico}

\date{\today}

\begin{abstract}
\noindent \textbf{Abstract.} 
Zero energy eigenstates $\psi_0(\theta)$ of the twisted bilayer graphene Hamiltonian at the Dirac point show a high sensitivity  to the twist angle $\theta$ near the magic angles where the effective Fermi velocity vanishes. We use  information theory markers, like  fidelity-susceptibility and entanglement entropy of the reduced density matrix to the layer sector, to capture this quantum criticality of zero modes at magic twist angles.
\end{abstract}


\maketitle

\section{Introduction}

After its rediscovery, isolation and investigation  by Novoselov and Geim \cite{Novoselov2004Science306,Novoselov2005PNAS102}, graphene and other 2D (thin layer) atomic crystals have become  valuable and useful nanomaterials (see e.g. \cite{RevModPhys.81.109} for a review on the electronic properties of graphene). The theoretical identification of graphene as a 2D topological (quantum spin Hall) insulator by  \cite{KaneMele05} also opened a field of interesting technological applications like  spintronic,  magnetoelectronic and optoelectronic devices and dissipationless transistors (see \cite{RevModPhys.82.3045,RevModPhys.83.1057} for traditional reviews), like FETs based on Bernal stacked bilayer graphene \cite{Icking2024}. 

The range of new technological possibilities on quantum devices increases when we add a stacking degree of freedom by going from one layer to a multilayer arrangement. Different bilayer graphene stackings exist~\cite{ROZHKOV20161}, of which twisted bilayer graphene (TBG) has raised much expectation because the twist angle generates Moir\'{e} patterns and leads to the emergence of new electronic properties like the observation of unconventional superconducting behavior at some ``magic'' twist angles like $\theta\simeq 1.1^{\circ}$~\cite{Cao2018unconventional,Cao2018correlated}. These findings of Jarillo-Herrero and coworkers confirmed previous predictions of Bistritzer and  MacDonald~\cite{Bistritzer12233} that the amount of energy a free electron would require to tunnel between two graphene sheets radically changes at this angle. More precisely, 
Ref.~\cite{Bistritzer12233} predicts that the effective/renormalized Fermi velocity in TBG at the Dirac point oscillates with twist angle, vanishing at a series of magic angles which give rise to large Dirac-point density-of-states (DOS)  and to large counter-flow conductivities (see also~\cite{PhysRevB.87.121402} for further studies about optical conductivities of TBG). Actually, the electronic spectrum is strongly affected by the varying twist angle, which  brings van Hove singularities \cite{Li2010,PhysRevLett.109.196802,PhysRevB.93.035452} to lower energies, making them accessible to the electrons and creating these exotic physical phenomena. 
Refs.~\cite{Cao2018unconventional,Cao2018correlated}  later experimentally confirmed these predictions in  showing that, owing to strong interlayer coupling, the electronic band structure near zero Fermi energy becomes flat when the twist angle is close to the magic angle $\theta\simeq 1.1^{\circ}$, giving rise to an unconventional superconductivity behavior. Zero modes therefore play an important role and are topologically protected~\cite{PhysRevB.84.045436}. The topological nature of magic angles was established in~\cite{PhysRevLett.123.036401}.

The analytic investigation of TBG performed by Bistritzer and MacDonald  in~\cite{Bistritzer12233} (usually named as BM models) was further pursued in a series of papers \cite{PhysRevB.103.205411,PhysRevB.103.205412,PhysRevB.103.205413,PhysRevB.103.205414,PhysRevB.103.205415,PhysRevB.103.205416} by Bernevig and coworkers. Previous low-energy models of TBG exist, like \cite{PhysRevLett.99.256802,PhysRevB.86.155449} by Lopes dos Santos and coworkers, but BM models are valid not only for commensurate twist angles but for arbitrary $\theta$. Other fundamental continuum models for TBG accounting for  the vanishing of the Fermi velocity and the perfect flattening of the entire lowest band were proposed by~\cite{PhysRevLett.122.106405}.   The literature on TBG is enormous and it is impossible to account for it all here. In this paper we shall follow the BM-like approach treating $\theta$ as a continuous control parameter.

Quantum information science viewpoints are increasingly being introduced into condensed matter physics. Actually, the study of topological order and the related new quantum phases is actually a study of patterns of entanglement~\cite{Jiang2012,ZengEntangTPT19}. More and more general methods to measure entanglement entropy in a many-body systems are proposed in the literature (see e.g. \cite{PhysRevLett.109.020504} for one of them based on a quantum switch coupled to several copies of the original system). 
R\'enyi-Wherl entropies and inverse participation ratio localization measures have proven to identify topological phase transitions in silicene~\cite{silicene1,silicene3}, phosphorene~\cite{MRX,IJQC} and HgTe/CdTe quantum wells~\cite{CALIXTO2022128057,calixto2024IPRHgTe}. 
The concept of fidelity-susceptibility has been quite useful in describing  traditional quantum phase transitions \cite{Zanardi,PhysRevLett.99.100603,GuPRE2007,GuIJMPB2010,octaviofide,octaviofide1,octaviofide2}. Here we want to explore this information tool in capturing some criticality at magic angles. Entanglement spectrum and lattice-layer entanglement has also been studied in~\cite{PhysRevB.93.115106} and \cite{PhysRevB.95.195145} for the particular case of Bernal-stacked bilayer graphene. Here we extend these studies to the more general TBG case, recovering the previous as a particular case, and demonstrating how magic angle signatures are printed in these quantities.

The organization of the paper is as follows. In Sec. \ref{modelsec} we review  all the essential  ingredients to construct the low-energy Hamiltonian model in a Hilbert-space basis conveniently adapted to a tensor product structure to deal with a composite Hilbert space. We make plots of the TBG Hamiltonian eigenspectrum and DOS for different twist angles, showing how low-energy flat moir\'e Bloch bands develop at magic angles. In Sec. \ref{Fermivelsec} we remind how flattening of moir\'e bands at magic angles is accompanied by a vanishing of the effective Dirac-point Fermi velocity. In Sec. \ref{secinfo} we provide information theory markers of magic twist angles as an alternative to the usual Fermi velocity. We show that sudden drops of fidelity for the lowest-energy (flat band) conduction and valence Hamiltonian eigenstates occur at exactly the magic angles. Resonances in the interlayer hopping/transfer probabilities (off diagonal components of the reduced density matrix  to the layer sector) and in the entanglement entropy also capture the magic angles. Finally, Sec. \ref{conclusec} is devoted to conclusions and outlook.

\section{Low-energy continuum model}\label{modelsec}

As already said, in order to describe the low-energy electronic properties of TBG, several Hamiltonian models have been developed in which Dirac electrons move in each layer and are hybridized by interlayer hopping. We shall also discard interaction between fermions. We understand that the reader is familiar with moir\'e patterns, superlattices and Brillouin zones (BZ) and we shall  give here just the essential ingredients to define the model Hamiltonian. 

\subsection{Momentum space, interlayer hopping and Hilbert space basis}

\begin{figure}
	\begin{center}
		\includegraphics[width=\columnwidth]{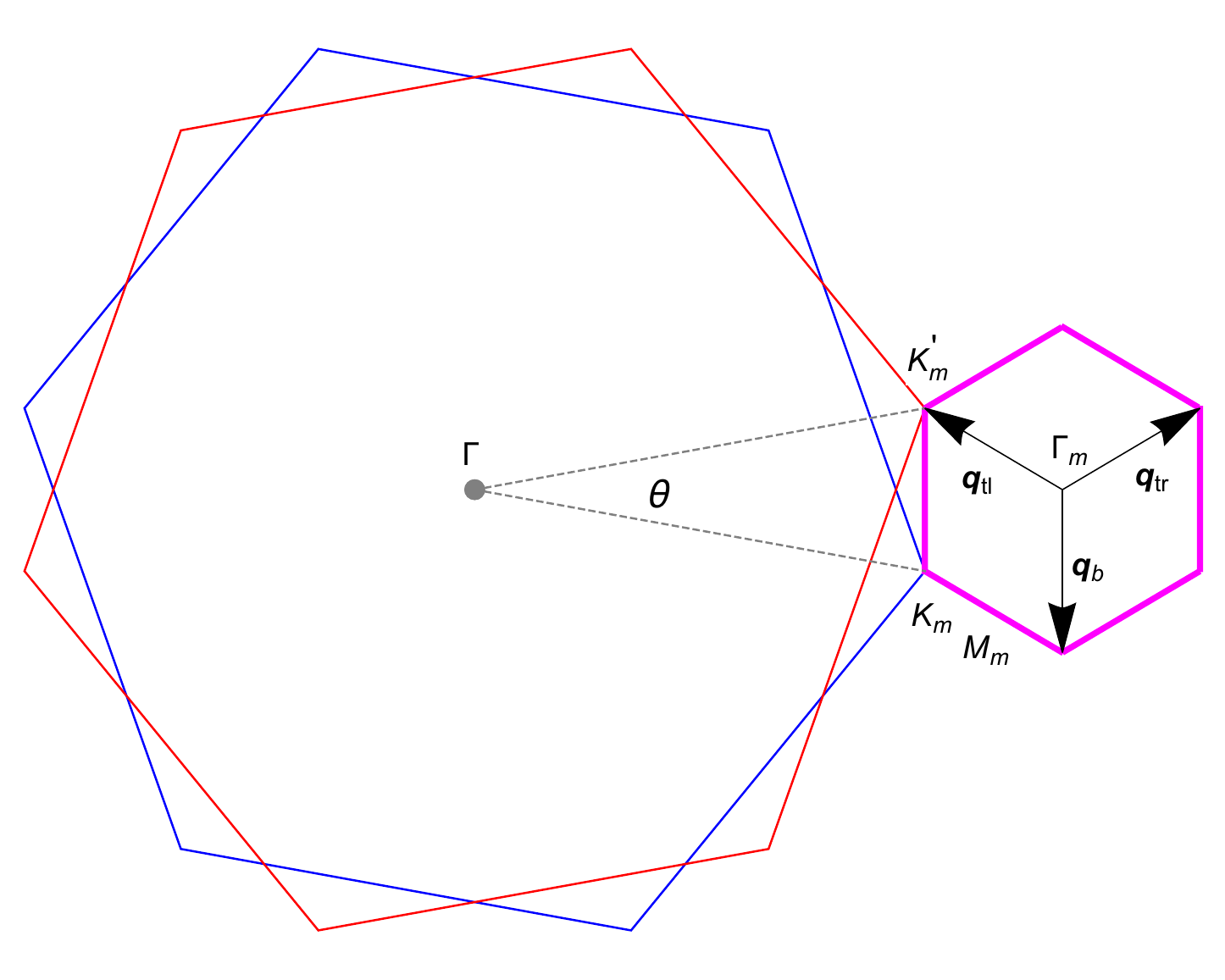}
	\end{center}
	\caption{Blue and red hexagons centered at $\Gamma$ describe the Brillouin zones for layers 1 and 2, respectively, which are rotated a twist angle $\theta$ to each other. The purple hexagon represents the moir\'e unit cell in reciprocal space for an expansion around the Dirac point $\bm{K}=\frac{4  \pi}{3  \sqrt{3}  d}(1,  0)$, with $d$ the carbon-carbon distance. The trajectory $K_m\to K_m'\to \Gamma_m\to M_m\to K_m$ (later used in Figs. \ref{fig1} and \ref{figEntropyTrayectory}) and the momentum transfer vectors $\bm{q}_{tl}, \bm{q}_{tr}, \bm{q}_{b}$ are shown together with the moir\'e  BZ.}
	\label{figBZ}
\end{figure}

The momentum transfer vectors of the three distinct inter-layer hopping processes [top-left ($tl$),  top-right ($tr$) and bottom ($b$)] are:  
\begin{eqnarray}
& \bm{q}_{tl} = K_\theta  (\frac{-\sqrt{3}}{2},\frac{1}{2}),\, \bm{q}_{tr} = K_\theta(\frac{\sqrt{3}}{2},\frac{1}{2}),
 & \nn\\ 
&\bm{q}_b = K_\theta  (0,-1),&
\end{eqnarray}
with $K_\theta=2K_D\sin{(\theta/2})$, where $K_D=\frac{4  \pi}{3  \sqrt{3}  d}$ is the modulus of the Dirac point vectors $\pm\bm{K}=\pm K_D (1,  0)$ and $d=0.142$~nm is the carbon-carbon distance in single-layer graphene. $K_\theta$ is the separation between the Dirac points of the two layers and establishes the size of the moir\'e BZ in the reciprocal space; see Fig. \ref{figBZ} for a graphical representation. These vectors are associated to the first three nearest-neighbors (NN) in reciprocal lattice between layers $\ell=1,2$. We can account for higher NN by defining (see \cite{Catarina2019,Beppe}) 
\begin{equation}
\bm{q}_{\bm{m}}^\ell=(\delta_{\ell,2}+m_1-m_2)\bm{q}_b-m_1\bm{q}_{tl}+m_2\bm{q}_{tr}, \label{waveNN}
\end{equation}
where the  two integers $(m_1,m_2)=\bm{m}$ determine the NN for each layer $\ell=1,2$ of graphene.

We shall simply denote  the set of basis states by:  
\begin{equation}
 |\ell,\bm{m},s\rangle\equiv|\ell,\bm{q}+\bm{q}_{\bm{m}}^\ell,s\rangle,\label{basisstates}
\end{equation}
where $s=A,B$ makes reference to the two triangular graphene sublattices and we just consider one single value of the spin third component (either up or down); we are measuring momentum $\bm{q}$ from the Dirac point $\bm{K}$ (similar results are obtained at $\bm{K}'=-\bm{K}$). These states form an orthonormal basis $\langle \ell,\bm{m},s|\ell',\bm{m}',s'\rangle=\delta_{\bm{m}\bm{m}'}\delta_{\ell\ell'}\delta_{ss'}$, where the Kronecker $\delta_{\bm{m}\bm{m}'}$ is understood on a component-by-component basis.

\begin{figure}
	\begin{center}
		\includegraphics[width=0.49\columnwidth]{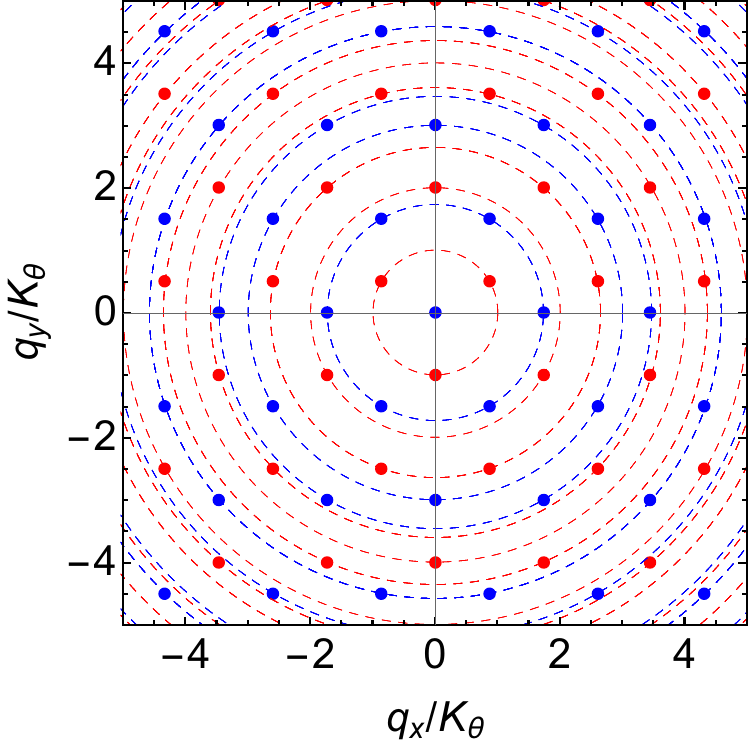}
		\includegraphics[width=0.49\columnwidth]{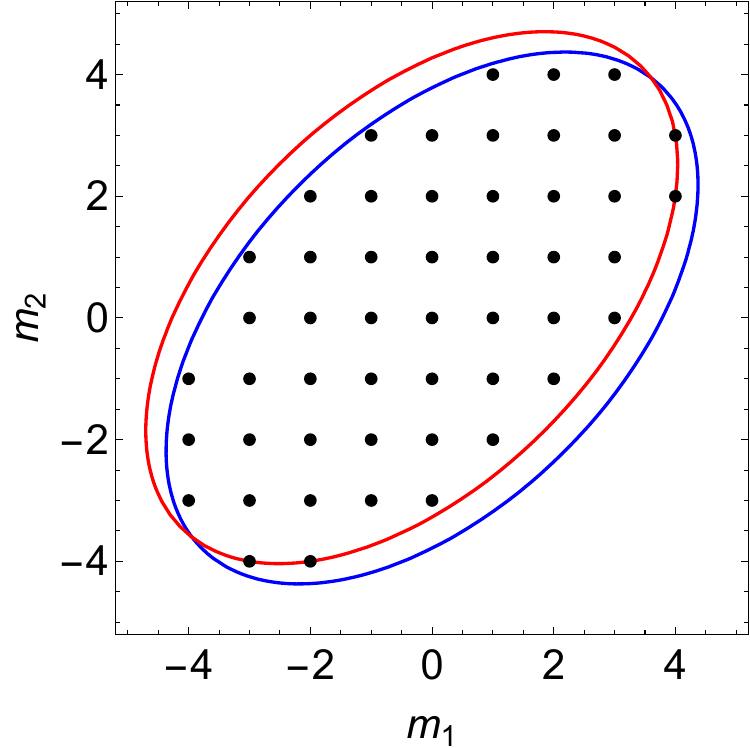}
	\end{center}
	\caption{Left panel:  wave vectors $\bm{q}_{\bm{m}}^\ell$ of Eq. \eqref{waveNN} (in $K_\theta$ units) are plotted as blue (layer $\ell=1$) and red (layer $\ell=2$) dots. We arrange them in circles of radius $R=|\bm{q}_{\bm{m}}^\ell|/K_\theta=1, \sqrt{3}, 2, \sqrt{7}, 3,  \dots,$ which yields Hamiltonian matrix orders of: $D=8, 20, 26, 38, 50, \dots$ Right panel: ellipses ${|\bm{q}_{\bm{m}}^\ell|}/{K_\theta}=\sqrt{43}$   are drawn in the plane $(m_1,m_2)$; common points to both ellipses $\ell=1,2$ are plotted in black color.
	}
	\label{fig0}
\end{figure}

By truncating the basis, we obtain an approximation for the electronic band structure in the moir\'e BZ. The size of the truncated Hamiltonian matrix depends on the number of higher-order NN chosen, which in turns depends on the modulus of the wave vector
\begin{equation}
\frac{|\bm{q}_{\bm{m}}^\ell|}{K_\theta}= \sqrt{(\ell-1)^2+3 ((\ell-1-m_2) (m_1-m_2)+m_1^2)}.
\end{equation}
The quantity $({|\bm{q}_{\bm{m}}^1|}/{K_\theta})^2/3=m_1^2-m_1m_2+m_2^2$ appears in the literature as \emph{L\"oschian numbers} (the norm of the \emph{Eisenstein integers}), with interesting properties in number theory. Note that $m_1^2-m_1m_2+m_2^2=n$  defines ellipses in the $(m_1,m_2)$ plane (see  Fig.~\ref{fig0}, right panel). In Figure \ref{fig0} (left panel) we draw wave vectors $\bm{q}_{\bm{m}}^\ell$ in $K_\theta$ units as blue dots for layer $\ell=1$ and red dots for layer $\ell=2$. This allows us to arrange the spinor basis states $|\ell,\bm{m},s\rangle$ according to the distance $R_\ell={|\bm{q}_{\bm{m}}^\ell|}/{K_\theta}$ to the origin of the moir\'e reciprocal lattice.  The Hamiltonian matrix size $D$ coincides with twice (due to the sublattice $s$ degree of freedom) the number of dots enclosed inside the circle of radius $R_\ell$. The minimum Hamiltonian matrix size corresponding to the first three NN of layer $\ell=2$, $(m_1,m_2)=(0,0), (0,1)$ and $(-1,0)$ (represented by red dots on the smallest red circle of radius $R_2=1$ in Fig. \ref{fig0}) is $D=8$; an explicit expression of the matrix Hamiltonian for this case is given in~\cite{Bistritzer12233}. For ten moir\'e reciprocal lattice vectors (represented by 3 red  and 7 blue dots inside the smallest blue circle of radius $R_1=\sqrt{3}$ in Fig. \ref{fig0}) one obtains a  Hamiltonian matrix size $D=20$, and so on. In general, the Hilbert space dimension $D$ is roughly proportional to the area of the circle of radius $R$. A linear fit of the first 109 moir\'e reciprocal lattice vectors inside a radius $R=\sqrt{43}$  gives
\begin{equation}
D\simeq 4.66 +  1.56\pi R^2.\label{DR}
\end{equation}

For our purposes, it will be more convenient to have a basis $\{|\ell,\bm{m},s\ra\}$ with a balanced number of basis vectors associated to each layer; for example, when computing reduced density matrices to the layer sector later in Sec. \ref{secinfo}, it is more convenient to adopt a tensor product structure
\begin{equation}
 |\ell,\bm{m},s\ra=|\ell\ra\otimes|\bm{m}\ra\otimes|s\ra.\label{basisstates2}
\end{equation}
Therefore, we shall chose moir\'e reciprocal lattice vectors $\bm{m}$ as common dots (black dots in Fig. \ref{fig0}, right panel) inside the two ellipses  ${|\bm{q}_{\bm{m}}^\ell|}/{K_\theta}=R_\ell\leq R, \ell=1,2$ in the plane $(m_1,m_2)$, for a given $R$. The formula \eqref{DR} still gives an upperbound for the corresponding Hamiltonian matrix size $D$. Actually, a new linear fit of $D$ as a function of $R^2\in[3,300]$ gives 
\begin{equation}
 D\simeq -11.94+4.69 R^2.
\end{equation}

Other more crude approaches would correspond to choosing points such as $\{(m_1,m_2): |m_1|, |m_2|\leq M\}$ (within squares) or $\{(m_1,m_2): |m_1+m_2|\leq M, |m_1-m_2|\leq M\}$ (inside rhombus) for a given threshold $M$.

\subsection{Hamiltonian matrix and electronic spectrum}

Let us use the shorthand convenient notation $|\ell,\bm{m}\ra$  for the set of basis states \eqref{basisstates2} when writing expressions in a  $2\times 2$ block-matrix form making reference to the two triangular sublattices $s=A, B$ of graphene. 
We shall rotate layer $\ell=1$ by $-\theta/2$ and layer $\ell=2$ by $\theta/2$. 
The twisted bilayer Hamiltonian matrix elements (in $2\times 2$ block-matrix form) at the Dirac point $\bm{K}$ are given by
\begin{eqnarray}
\langle \bm{m},\ell|H|\bm{m}',\ell\rangle&=&  \mathbb{h}(\bm{q}+\bm{q}_{\bm{m}}^\ell,(-1)^\ell\theta/2)\delta_{\bm{m},\bm{m}'},\nn\\
\langle \bm{m},1|H|\bm{m}',2\rangle&=& \mathcal{T}_{\bm{q}_{tl}}\delta_{\bm{m},\bm{m}'+(1,0)}+\mathcal{T}_{\bm{q}_{tr}}\delta_{\bm{m},\bm{m}'-(0,1)}\nn\\ &&+\mathcal{T}_{\bm{q}_{b}}\delta_{\bm{m},\bm{m}'},\label{HTBG}
\end{eqnarray}
where 
\begin{equation}
 \mathbb{h}(\bm{q},\theta)=\hbar v |\bm{q}|
 \begin{pmatrix} 0& e^{-\ic(\theta_{\bm{q}}+\theta)}\\ e^{ \ic(\theta_{\bm{q}}+\theta)} & 0\end{pmatrix},
\end{equation}
is the  $2\times 2$ Dirac Hamiltonian for the  single layer graphene (we chose Fermi velocity $v$ so that $\hbar v=583$~meV.nm as in Ref.~\cite{Cao2018correlated}) rotated an angle $\theta$, and $\bm{q}$ (with polar components $|\bm{q}|$ and $\theta_{\bm{q}}$) is measured from the  Dirac point $\bm{K}$, assuming that the deviations are small compared to the BZ dimensions. 
The interlayer hopping interaction matrices  are given by

\begin{eqnarray}
& \mathcal{T}_{\bm{q}_{tl}} = w
\begin{pmatrix}
e^{-\ic  \phi}  &   1  \\
e^{\ic \phi}  &   e^{-\ic \phi} 
\end{pmatrix},\; 
\mathcal{T}_{\bm{q}_{tr}} = w 
\begin{pmatrix}
e^{\ic \phi}  &   1  \\
e^{-\ic \phi}  &   e^{\ic \phi} 
\end{pmatrix} ,  &\nn\\ &\mathcal{T}_{\bm{q}_b}=  w 
\begin{pmatrix}
1  &   1  \\
1  &   1 
\end{pmatrix},&
\end{eqnarray}
with $w=110$~meV the hopping energy, and $\phi=2\pi/3$.

\begin{figure}
	\begin{center}
		\includegraphics[width=4.2cm]{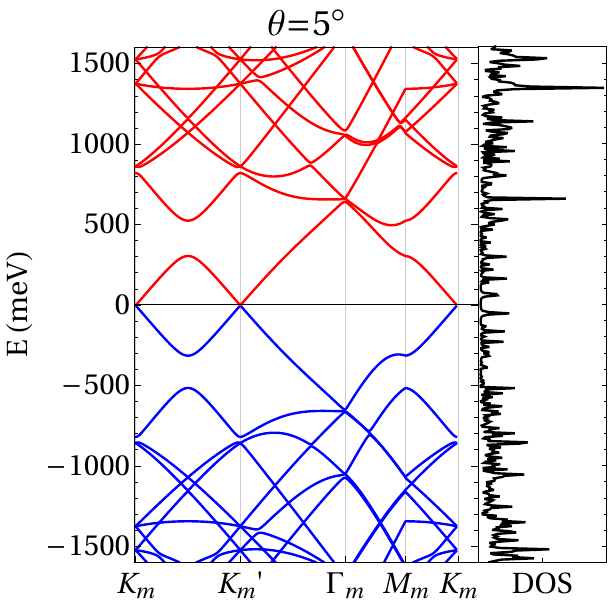}
		\includegraphics[width=4.2cm]{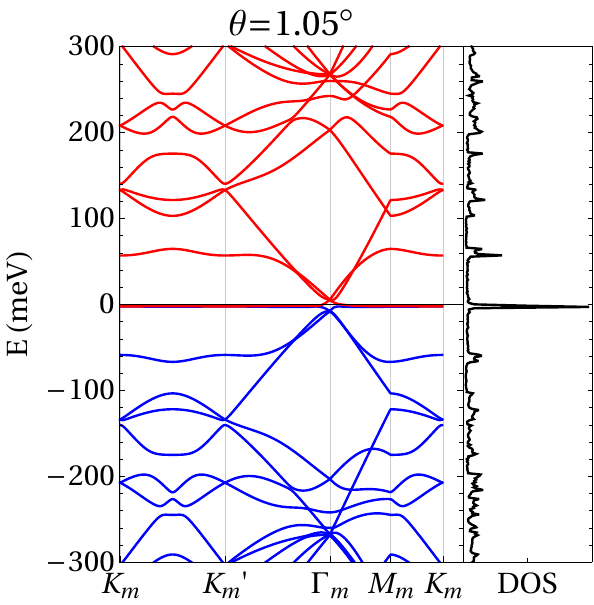}\\[2mm]
		\includegraphics[width=4.2cm]{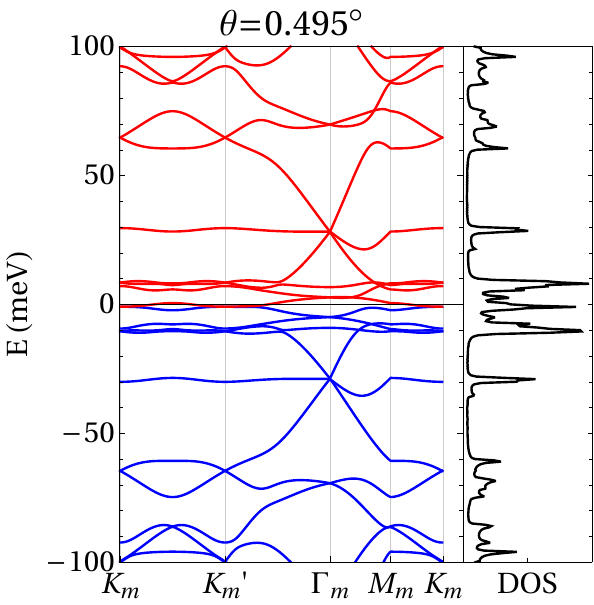}
		\includegraphics[width=4.2cm]{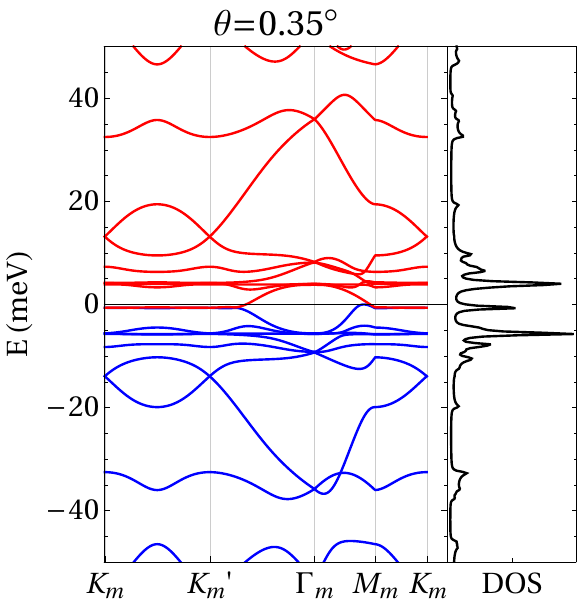}\\[2mm]
		\includegraphics[width=4.2cm]{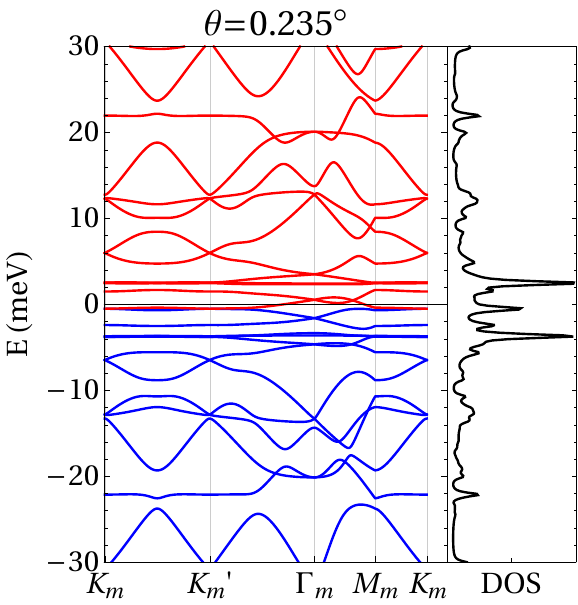}
		\includegraphics[width=4.2cm]{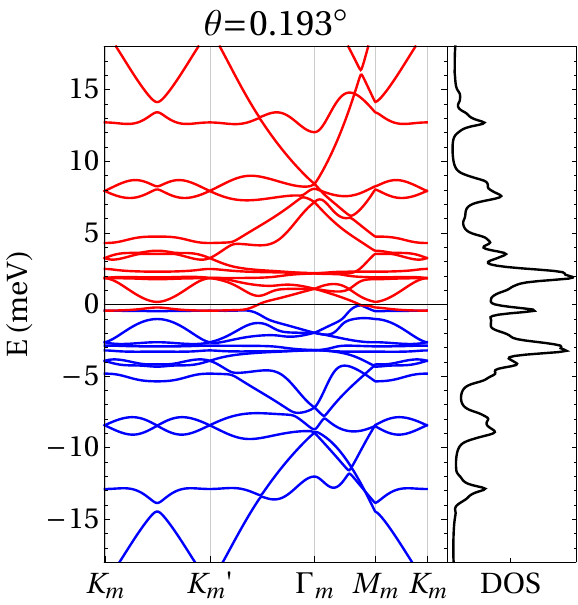}
	\end{center}
	\caption{TBG band structure and DOS as a function of the BZ momentum $\bm{q}$ along the trajectory $K_m\to K'_m\to \Gamma_m\to M_m\to K_m$ of Fig. \ref{figBZ}  for different twist angles.  Low-energy flat bands develop at magic angles $\theta\simeq 1.05^\circ, 0.495^\circ, 0.35^\circ, 0.235^\circ$ and $ 0.193^\circ$, as seen in the corresponding DOS. 
	}
	\label{fig1}
\end{figure}

In Figure~\ref{fig1}  we represent the TBG Hamiltonian eigenspectrum  (energy dispersion) and DOS as a function of the BZ momentum $\bm{q}$  along the trajectory $K_m\to K'_m\to \Gamma_m\to M_m\to K_m$ of Fig. \ref{figBZ}  for different twist angles; five of them are magic angles later captured by the Fermi effective velocity and information-theoretic measures (see Fig \ref{figFermiFidelity}).  The low energy spectrum is similar to that of isolated graphene for large twist angles (we show the case $\theta=5^\circ$). However, as the twist angle decreases, the number of bands in a given energy window near zero increases. This can also be explained by an appearance of  van Hove singularities at zero energy where $\nabla_{\bm{q}}E_0(\bm{q})\simeq \bm{0}$. Low-energy flat moir\'e Bloch bands develop at magic angles, which contribute to sharp peaks in the corresponding DOS measured as
\be
D_\eta(\omega)=\sum_E \delta_\eta(\omega-E).
\ee
Here we are using the Lorentzian $\delta_\eta(x)=(\eta/\pi)/(\eta^2+x^2)\to\delta(x)$ as a finite representation/regularization of the Dirac delta for small $\eta$. We take $\eta=0.25$~meV.

We expect convergence by truncating momentum space at lattice vectors of the order of $k_0\sim w/(\hbar v)$. This means $R\sim k_0/K_\theta$ for a given angle $\theta$. To be on the safe side, we chose around five times this quantity, which leads to  $R\simeq \sqrt{10}/\theta$ for small angles $\theta$  in degrees. For example, to capture with convergence the smallest magic angle $\theta_5=0.193^\circ$ we have chosen $R=\sqrt{300}$, which means a Hamiltonian matrix size of $D=1388$. We shall use $D=1388$ all along the manuscript unless otherwise stated.

\section{Effective Fermi velocity, flat bands, zero modes and magic angles}\label{Fermivelsec}

\begin{figure}
	\begin{center}
		\includegraphics[width=\columnwidth]{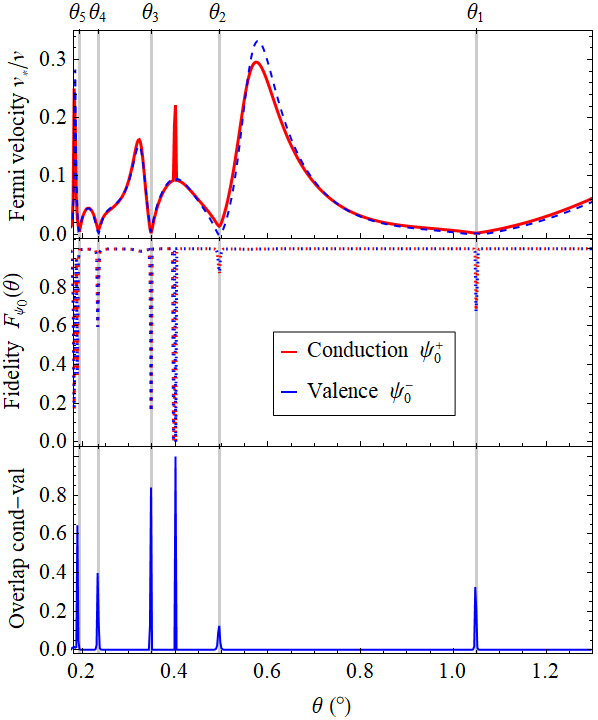}
	\end{center}
	\caption{Top panel: Effective Fermi velocity $v_*/v$ \eqref{FermiV}  in the proximity of the Dirac point $\bm{K}_m$ as a function of the twist angle $\theta$ (in degrees) for the lowest-energy (flat band) conduction (red) and valence (blue)  Hamiltonian eigenstates $\psi_0^\pm$. Fermi velocity vanishes at the magic angles: $\theta_1=1.05^\circ$ (principal),  $\theta_2=0.495^\circ, \theta_3=0.35^\circ, \theta_4=0.235^\circ, \theta_5=0.193^\circ$.  Middle panel: Fidelity $|\langle \psi_{0}^\pm(\theta) |\psi_{0}^\pm(\theta+\delta\theta)\rangle|^2$ [conduction (+) in red and valence (-) in blue] as a function of $\theta$; sudden fidelity drops occur at magic angles. Bottom panel: Fidelity $|\langle \psi_{0}^+(\theta) |\psi_{0}^-(\theta+\delta\theta)\rangle|^2$ between the lowest energy conduction and valence Hamiltonian eigenstates, displaced a quantity $\delta\theta$ relative to each other,  as a function of $\theta$;  sudden fidelity rises occur at magic angles, and also, at an additional ``anomalous  angle'' $\theta\simeq 0.4^\circ$ (explained in the main text). 
	}\label{figFermiFidelity}
\end{figure}

 Flattening of moir\'e bands at magic angles is accompanied by a vanishing of the effective Dirac-point Fermi velocity $\bm{V}_*=\frac{1}{\hbar}\left.\nabla_{\bm{q}}H\right|_{\bm{q}=\bm{0}}$. Indeed, in Figure \ref{figFermiFidelity} (top panel) we represent the renormalized effective Fermi velocity expectation value (in Fermi velocity $v$ units)
\begin{equation}
v_*/v=|\langle \psi_0^\pm|\bm{V}_*|\psi_0^\pm\rangle|/v\label{FermiV}
\end{equation}
 in a quasi-zero energy $E_0^\pm$ normalized eigenstate $\psi_0^\pm$ [conduction (+) and valence (-)] of the Hamiltonian \eqref{HTBG} at $\bm{q}=\bm{0}$, as a function of the twist angle $\theta$. Conduction and valence eigenstates $\psi_0^\pm$ turn out to give roughly the same effective Dirac-point  velocity $v_*$ (see red and blue curves in Fig. \ref{figFermiFidelity}, top panel), which vanishes at magic angles $\theta\simeq 1.05^\circ, 0.495^\circ, 0.35^\circ, 0.235^\circ$ and $ 0.193^\circ$. These magic angles are similar to the ones obtained in the literature (see e.g. \cite{Bistritzer12233}). Note that their particular values are sensitive to the choice Hamiltonian parameters like the  Fermi velocity $v$ and the hopping energy $w$.

\begin{figure}
	\begin{center}
		\includegraphics[width=\columnwidth]{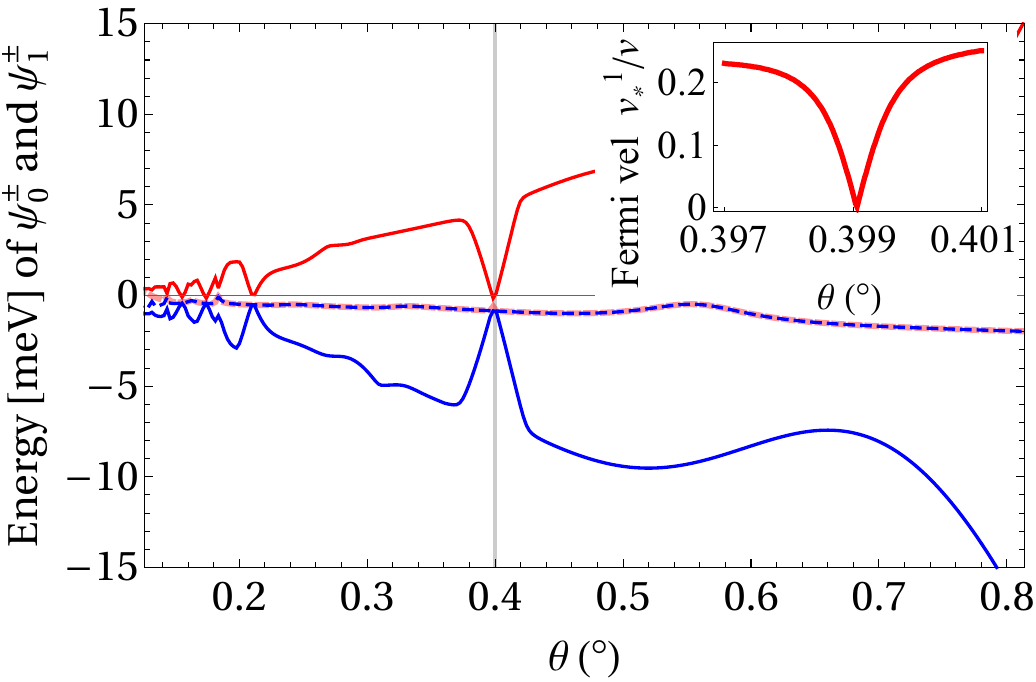}
		\end{center}
	\caption{Energies of quasi-zero $\psi_0^\pm$ (thickest curves)  and first excited $\psi_1^\pm$ Hamiltonian eigenstates [conduction (+) in red and valence (-) in blue] as a function of the twist angle $\theta$ (in degrees). Energies $E_0^\pm$ and $E_1^\pm$ degenerate at the twist angle $\theta\simeq 0.4^\circ$ (marked with a grey vertical  gridline). This degeneracy seems to be responsible for the sudden drop of fidelity and the outlier value   of the Fermi velocity  \eqref{FermiV}  at $\theta\simeq 0.4^\circ$, between the magic angles $\theta_2$ and $\theta_3$  in Fig. \ref{figFermiFidelity}. The inset shows that $\theta\simeq 0.4^\circ$ actually corresponds to an excited magic angle where the Fermi velocity expectation value 
$v_*^1$ in Eq.   \eqref{FermiV1}  vanishes for the first excited conduction state $\psi_1^+$.
	}
	\label{figexcited}
\end{figure}

 There is an outlier value of the effective Fermi velocity at the twist angle $\theta\simeq 0.4^\circ$,  between the magic angles $\theta_2$ and $\theta_3$  in the top panel of Fig. \ref{figFermiFidelity}, perceived as an anomalous pick when joining the points of the list plot. This outlier value persists for other Hamiltonian model parameters, like  $v$ or  $w$,  and matrix sizes $D$, although its location varies with $v$ and $w$, so do the magic angles. Therefore, an explanation is necessary for this outlier. It turns out to be related  to a degeneracy $E_0^\pm\simeq E_1^\pm$ between the lowest (quasi-zero) energy  $\psi_0^\pm$ and the first excited $\psi_1^\pm$ Hamiltonian eigenstates at  $\theta\simeq 0.4^\circ$, as can be seen in Fig.~\ref{figexcited}. Moreover, $\theta\simeq 0.4^\circ$ turns out to be an ``excited magic angle" in the sense that the effective Fermi velocity expectation value 
 \begin{equation}
 v_*^1/v=|\langle \psi_1^\pm|\bm{V}_*|\psi_1^\pm\rangle|/v \label{FermiV1}
 \end{equation}
 in the first excited states $\psi_1^\pm$ vanishes (see the inset in Fig \ref{figexcited}). This degeneracy between $\psi_0^\pm$ and  $\psi_1^\pm$ at $\theta\simeq 0.4^\circ$ also manifest itself in a sudden fidelity drop perceived in the middle panel of Fig. \ref{figFermiFidelity} (see Sec. \ref{fidsec} for more details). We have also revealed the existence of other ``excited magic angles'', like $\theta\simeq 0.52, 0.66^\circ$, where  $v_*^1$ vanishes, although this degeneration phenomenon was not observed and therefore they do not manifest in the zero-mode effective Fermi velocity \eqref{FermiV}. The possible relevance and  physical significance of these excited magic angles is yet to be determined.

 In the following sections we shall provide alternative markers of magic angles from a quantum information theory perspective.

\section{Information theory markers of magic twist angles}\label{secinfo}

We are going to study two information measures like fidelity and entanglement entropy and see how they also are able to capture magic angles in TBG.

\subsection{Fidelity-susceptibility}\label{fidsec}

Fidelity is an elemental concept in general information theory. It  measures the accuracy of transmission. In quantum theory, the scalar product $F=|\la\psi_1|\psi_2\ra|^2$ gives the fidelity (closeness) between two normalized pure states $|\psi_1\ra$ and $|\psi_2\ra$. Fidelity turns out to be a good marker/precursor of a quantum phase transition even  for finite size systems near a critical point (see for example 
\cite{Zanardi,PhysRevLett.99.100603,GuPRE2007,GuIJMPB2010,octaviofide,octaviofide1,octaviofide2}).

Here we propose this measure as a  marker of magic angles in TBG, where sudden changes in Hamiltonian eigenstates take place. We will  study the fidelity of lowest (quasi-zero) energy Hamiltonian eigenstates $\psi_0^\pm$  at the Dirac point between  $\theta$ and $\theta
+ \delta \theta$: 
\be
F_{\psi_0^\pm}(\theta)=|\langle \psi_0^\pm(\theta)|\psi_0^\pm(\theta+\delta\theta)\rangle|^2,\label{fidelityeq}
\ee
as a function of $\theta$. We expect sudden drops of $F_{\psi_0^\pm}(\theta)$ at magic (critical) twist angles $\theta_c$ if $\psi_0^\pm$ suffers  drastic structural changes at $\theta_c$, depending on $\delta\theta$. A higher order approximation which is less sensitive to the choice of $\delta\theta$ is the \emph{fidelity susceptibility} 
\be
\chi_{\psi_0}(\theta)= 2\frac{1-F_{\psi_0}(\theta,\theta
+ \delta \theta)}{\delta\theta^2}.\label{susceptibilityeq}
\ee

In Figure \ref{figFermiFidelity} (middle panel) we plot the fidelity $F_{\psi_0^\pm}$ of the lowest energy (quasi-zero) $E_0^\pm$ Hamiltonian eigenstates $\psi_0^\pm$  at the Dirac point $\bm{q}=\bm{0}$ as a function of $\theta\in[0.18,1.3]$ in degrees (we take a step $\delta\theta=0.0023^\circ$). We identify sudden drops of fidelity exactly at the magic angles $\theta_1=1.05^\circ$ (principal),  $\theta_2=0.495^\circ, \theta_3=0.35^\circ, \theta_4=0.235^\circ$ and $\theta_5=0.193^\circ$ where the effective Fermi velocity $v_*$ drops to zero. 

Additionally, fidelity also detects the ``excited magic  angle''  $\theta\simeq 0.4^\circ$,  where the effective Fermi velocity $v_*^1$ in Eq. \eqref{FermiV1} vanishes. As already commented at the end of Sec. \ref{Fermivelsec}, the origin of this fidelity drop is linked to the fact that the quasi-zero energy $\psi_0^\pm$ and first excited $\psi_1^\pm$ Hamiltonian eigenstates almost degenerate at  $\theta\simeq 0.4^\circ$.   Fidelity is sensitive to this degeneracy since $\psi_0^\pm$ and  $\psi_1^\pm$ are orthogonal, as are $\psi_0^+$ and $\psi_0^-$.

For completeness, Figure~\ref{figFermiFidelity} (bottom panel) shows how  magic angles are also detected by sudden rises of the overlap
\begin{equation}
 |\langle \psi_{0}^+(\theta) |\psi_{0}^-(\theta+\delta\theta)\rangle|^2
\end{equation}
between lowest (quasi-zero) energy conduction and valence Hamiltonian eigenstates displaced a quantity $\delta\theta$ relative to each other. This means that $\psi_{0}^\pm(\theta)$ and $\psi_{0}^\mp(\theta+\delta\theta)$ become alike at the magic angles $\theta_i, i=1,\dots,5$.

\subsection{Reduced density matrix and entanglement measures}\label{secentang}

We now study entanglement measures of the reduced density matrix (RDM) to the layer sector of low energy Hamiltonian eigenstates $\psi_{0}^\pm$. We firstly consider the simplest case of zero twist angle  (Bernal stacking in particular) as a first step towards the more general case. Entanglement spectrum for Bernal stacking has also been considered before in the literature like in  \cite{PhysRevB.93.115106} and  \cite{PhysRevB.95.195145}.

\subsubsection{Bernal stacking}

Bilayer graphene in a Bernal-stacked form is well described by the  $4\times 4$ Hamiltonian (at the Dirac points $\pm\bm{K}$)
\begin{equation}
H_B= \hbar v \left(
\begin{array}{cccc}
 0 & |\bm{q}|e^{\mp \ic\theta_{\bm{q}}} & 0 & \frac{t_\perp}{\hbar v} \\
 |\bm{q}|e^{\pm \ic\theta_{\bm{q}}} & 0 & 0 & 0 \\
 0 & 0 & 0 & |\bm{q}|e^{\mp \ic\theta_{\bm{q}}} \\
 \frac{t_\perp}{\hbar v} & 0 & |\bm{q}|e^{\pm \ic\theta_{\bm{q}}}  & 0 \\
\end{array}
\right),
\end{equation}
where   $t_\perp=3w= 330$~meV is the interlayer coupling constant. 
The corresponding four Hamiltonian eigenvalues can be obtained in analytic form as
\begin{eqnarray}
E_{\pm 1}&=&\mp \frac{t_\perp}{2} ( 1 - \sqrt{1 + \eta^2}  ), \\  E_{\pm 2}&=&\pm \frac{t_\perp}{2} ( 1 + \sqrt{1 + \eta^2} ),\nn
\end{eqnarray}
where we have defined the dimensionless parameter $\eta = {2  |\bm{q}|  \hbar v}/{t_\perp}=3.83 |\bm{q}|$ (with $|\bm{q}|$ is measured in nm${}^{-1}$). Notice that $|E_{\pm 1}|<|E_{\pm 2}|$, where $\pm$ makes reference to valence ($-$) and conduction ($+$) bands, as we have been reporting so far. The corresponding four Hamiltonian eigenvectors are
\begin{eqnarray}
|\psi_{\pm 1}\rangle&=& (\pm c_2,  c_1 e^{i \theta_{\bm{q}}}, \pm c_1e^{-i \theta_{\bm{q}}}, c_2  )^T  \label{Hameigen}\\
|\psi_{\pm 2}\rangle&=&  (\mp c_1 , -c_2  e^{i \theta_{\bm{q}}}  , c_2 e^{-i \theta_{\bm{q}}} , c_1 )^T , \nonumber 
\end{eqnarray}
with coefficients
\begin{equation}
c_1 = \frac{1}{2} \sqrt{1 + \frac{1}{\sqrt{1 +\eta^2}}} \, , \quad c_2 =  \frac{\eta }{2 \left(1+\eta^2 +\sqrt{1 + \eta^2} \right)^{\frac{1}{2}} }.
\end{equation}
We are using  the basis vectors 
\begin{equation}
 \left\{|\ell\rangle\otimes|s\rangle, \; \ell=1,2,\,  s=A,B\right\},
\end{equation}
like a two-qubit system with $\ell=1,2$ (resp. $s=A,B$) making reference to the layer sector (resp. sublattice sector). Therefore we have a composite Hilbert space $\mathcal{H}_\ell\otimes \mathcal{H}_s$. One can construct the $4 \times4$ density matrices $\rho_{\pm n}=|\psi_{\pm n}\rangle\langle \psi_{\pm n}|$ of the corresponding four Hamiltonian eigenstates \eqref{Hameigen}. If we want to focus on the layer sector $\mathcal{H}_\ell$, we can calculate the corresponding RDM  $\varrho_{\pm n}=\tr_2(\rho_{\pm n})$ by tracing over $\mathcal{H}_s$. A straightforward calculation gives 
\begin{eqnarray}
\varrho_{\pm n} = \begin{pmatrix}
  \frac{1}{2} &   \frac{(-1)^{n+1}\eta e^{i \theta_{\bm{q}}} }{2 \sqrt{ 1 + \eta^2}}  \\
 \frac{(-1)^{n+1}\eta e^{-i \theta_{\bm{q}}} }{2 \sqrt{ 1 + \eta^2}} &   \frac{1}{2}
\end{pmatrix},\; n=1,2,
\end{eqnarray}
with identical results for valence and conduction bands. 
The diagonal RDM  components ${P}_{\ell}=(\varrho_n)_{\ell\ell}=1/2$ give information about the probability to find the electron at layers $\ell=1,2$ (that is,  $50\%$), whereas the modulus of off-diagonal RDM component ${T}=|(\varrho_n)_{12}|=\eta/(2\sqrt{1+\eta^2})$ gives information about the transfer/hopping probability between layers (that is, ${T}=0$ at the Dirac point $\bm{q}=\bm{0}$ and ${T}\to 1/2$ far from the Dirac point). These quantities turn out to be independent of the particular  Hamiltonian eigenvector $\psi_{\pm n}, n=1,2$.

In order to see if $\psi_{\pm n}$ are entangled or not, we calculate either the  linear $L_n=1-\tr(\varrho_{n}^2)$ or the  von 
Neumann $S_n=-\tr(\varrho_n\ln \varrho_n)$ entropies, which can be written in terms of the previously defined parameter $\eta$ as (they turn out to be independent of $n=1,2$) 
\begin{equation}
L= \frac{1}{2 ( 1 + \eta^2)}  , \quad S = -e_+ \log{e_+} -e_- \log{e_-}  ,
\end{equation}
with $e_\pm = 1/2 \left( 1 \pm \eta/\sqrt{1 + \eta^2}\right)$ denoting the eigenvalues of $\varrho_n$. From these expressions we conclude that maximum entanglement entropy (i.e., $L=1/2$)  is attained at the Dirac point $ |\bm{q}|=0 \rightarrow\eta=0$, whereas $\psi_{\pm n}$ are almost pure states (i.e. $L\simeq 0$) far from the Dirac point. 

These results do not seem to depend on the particular entanglement measure used. 
There are many more entanglement measures between two qubits in the literature. For example, the one  proposed by Schlienz \& Mahler \cite{Mahler,mahlerBook}, which is defined as follows. The $4\times 4$ density matrix $\rho$ is now written in terms of the 16 generators of the unitary group U(4), which can be written as tensor products of Pauli matrices $\bm{\sigma}=(\sigma_1,\sigma_2,\sigma_3)$ plus the $2\times 2$ identity matrix $\sigma_0$. More precisely
\begin{eqnarray}
\rho &=& \frac{1}{4}\sigma_0 \otimes \sigma_0+ \frac{1}{4} \sum_{k=1}^3 (\lambda^{(1)}_k \sigma_k \otimes \sigma_0+   \lambda^{(2)}_k \sigma_0 \otimes \sigma_k)  \nn\\ &&+  \frac{1}{4} \sum_{k, j} C^{(1,2)}_{kj} \sigma_k \otimes \sigma_{j} \, ,
\end{eqnarray}
with 
 \begin{eqnarray}
&\bm{\lambda}^{(1)} = \tr(\rho\,\bm{\sigma} \otimes \sigma_0) , \, \bm{\lambda}^{(2)} = \tr(\rho\,\sigma_0\otimes \bm{\sigma} ), &\nn\\ 
&{C}^{(1,2)} _{kj}= \tr(\rho\,{\sigma}_k \otimes {\sigma}_j),&
\end{eqnarray}
where $\bm{\lambda}^{(1)}$ and $\bm{\lambda}^{(2)}$ denote the Bloch coherence vector of the layer and sublattice sectors, respectively,  and the  $3 \times 3$ matrix ${C}^{(1,2)} $ accounts for correlations between both sectors. The RDM on the layer sector is then
\begin{equation}
\rho^{(1)}=\tr_2(\rho)=\frac{1}{2}\sigma_0+\frac{1}{2}\sum_{k=1}^3\lambda^{(1)}_k\sigma_k \otimes \sigma_0,
\end{equation}
and analogously on sublattice sector $\rho^{(2)}$. Comparing $\rho$ with the direct product $\rho^{(1)}\otimes \rho^{(2)}$, the difference comes from a $3\times 3$ entanglement matrix $M$ with components
\begin{equation}
{M}_{jk} = {C}^{(12)}_{jk} - \lambda^{(1)}_j \lambda^{(2)}_k ,\; j,k = 1,2,3.\label{MMahler}
\end{equation}
Based on $M$, authors in Ref. \cite{Mahler} introduce a measure of qubit-qubit entanglement given by the parameter 
\begin{equation}
\beta = \frac{1}{3} \tr({M}^T {M} ),
\end{equation}
which is bounded by $0\leq \beta \leq 1$ and carries  information about the correlations between layer and sublattice sectors. In our case, it  takes the same value
\begin{equation}
\beta= \frac{3 + 2 \eta^2}{3 (1 + \eta^2) ^2 } ,\label{BernalBeta}
\end{equation}
 for the density matrices of the four Hamiltonian eigenstates \eqref{Hameigen}.  Again, maximum entanglement ($\beta=1$) is attained at  $\eta=0\rightarrow |\bm{q}|=0$, whereas $\psi_{\pm n}$ are almost pure states ($\beta\to 0$) far from the Dirac point.
 
  Besides, by means of the positive partial transpose criterion of Peres-Horodecki \cite{PhysRevLett.77.1413,HORODECKI19961,RevModPhys.81.865-Horodecki}, one is able to demonstrate that the system is not separable because the partial transpose matrix of the density matrix has a negative eigenvalue.

\subsubsection{RDM and entanglement at magic angles}

Now we are in position to analyze the layer sector of TBG and see how the corresponding RDM components and entropies somehow capture magic angles.

Let us expand Hamiltonian \eqref{HTBG} eigenvectors $\psi$ in the basis \eqref{basisstates2} as
\begin{equation}
 |\psi\ra=\sum_{\ell=1}^2\sum_{\bmm}\sum_{s=A}^B c_{\bmm s}^\ell|\ell,\bmm,s\ra.
\end{equation}
The corresponding $D\times D$ density matrix acquires the form 
\begin{equation}
 \rho=|\psi\ra\la\psi|=\sum_{\ell,\ell'}\sum_{\bmm,\bmm'}\sum_{s,s'}\rho_{\bmm s,\bmm' s'}^{\ell\ell'}|\ell,\bmm,s\ra\la\ell',\bmm',s'|.
\end{equation}
The RDM to the layer sector is
\begin{equation}
 \varrho=\begin{pmatrix} 
          \varrho_{11}& \varrho_{12}\\ \varrho_{21} &\varrho_{22}
         \end{pmatrix}=
\sum_{\bmm}\sum_{s}\begin{pmatrix}  \rho_{\bmm s;\bmm,s}^{11} &\rho_{\bmm s,\bmm s}^{12}\\ \rho_{\bmm s,\bmm s}^{21}& \rho_{\bmm s,\bmm s}^{22}\end{pmatrix}.
\end{equation}
We remind that the diagonal entries ${P}_\ell=\varrho_{\ell\ell}$ represent the probabilities to find the electron at layers $\ell=1,2$. The non-diagonal ${T}=|\varrho_{12}|$ part represents the probability transfer between layers (interlayer hooping). The linear entropy  $L=1-\tr(\varrho^2)$ takes the maximum value $L_\mathrm{max.}=1/2$ for a maximally entangled state and $L_\mathrm{min.}=0$ for a pure state. 

\begin{figure}
	\begin{center}
		\includegraphics[width=8cm]{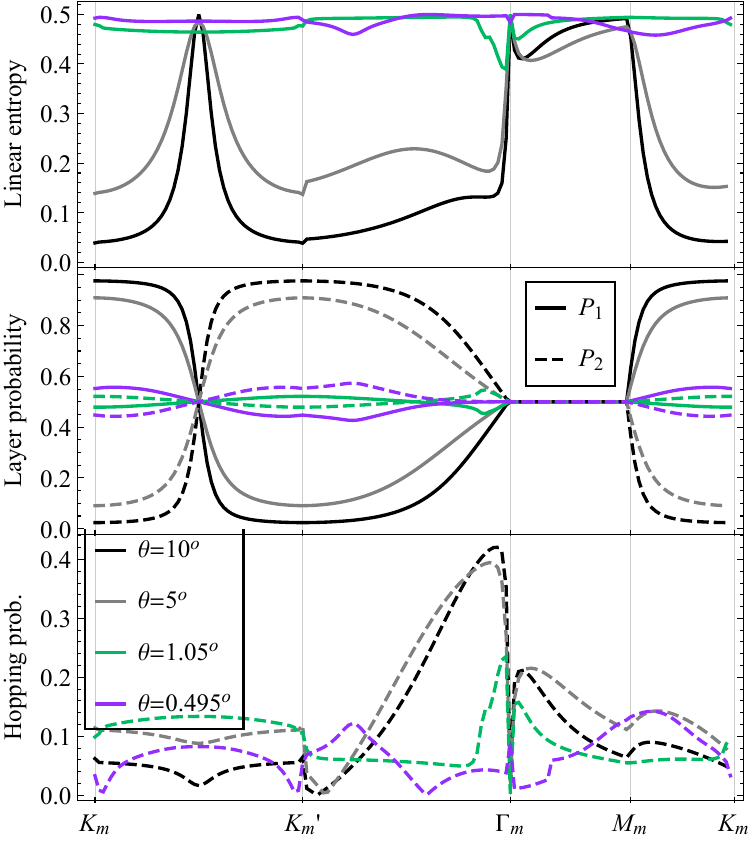}
	\end{center}
	\caption{Linear entropy $L$ (top panel), layer probabilities ${P}_\ell$ (middle panel), $\ell=1$ in solid and $\ell=2$ in dashed, and probability transfer ${T}$ between layers (bottom panel). All quantities are calculated for the lowest-energy valence Hamiltonian eigenstate $\psi_0^-$ as a function of the BZ momentum $\bm{q}$ 	
	along the trajectory $K_m\to K'_m\to \Gamma_m\to M_m\to K_m$ of Fig. \ref{figBZ}, and for different twist angles: $\theta=0.495^\circ, 1.05^\circ, 5^\circ$ and $10^\circ$.}
	\label{figEntropyTrayectory}
\end{figure}

In Fig. \ref{figEntropyTrayectory} we represent these quantities for the  valence Hamiltonian eigenstate $\psi_0^-$ (similar results are obtained for conduction $\psi_0^+$ eigenstate) as a function of the BZ momentum $\bm{q}$ along the trajectory $K_m\to K'_m\to \Gamma_m\to M_m\to K_m$ of Fig. \ref{figBZ}  for different twist angles (two of them magic). Just as in Fig.~\ref{fig1} we saw the development of flat energy bands at magic angles, we now see in Fig.~\ref{figEntropyTrayectory} (top panel) the emergence of ``flat'' (maximal $L\simeq 1/2$) entanglement entropy bands. We only show the two magic angles $\theta_1=1.05^\circ$ and $\theta_2=0.495^\circ$ as opposite to the two non magic angles $\theta=5, 10^\circ$ for which entropy oscillates between 0 and 1/2 along the given trajectory (the amplitude of the oscillations is greater for larger angles). Layer probabilities ${P}_\ell$ (middle panel) remain close to 1/2 for magic angles all along the trajectory, while they strongly oscillate  for large twist angles, showing sections in the trajectory where the electron prefers one layer over the other, and points where there are layer inversions; in the section $\Gamma_m\to M_m$ it happens that both layers are equally probable for all angles. Interlayer hopping ${T}$ (Fig.~\ref{figEntropyTrayectory} bottom panel) is perhaps not so easy to interpret, but we could say that there is a high electron transfer probability between layers for larger angles at middle points in section $K'_m\to \Gamma_m$.

\begin{figure}
	\begin{center}
\includegraphics[width=8cm]{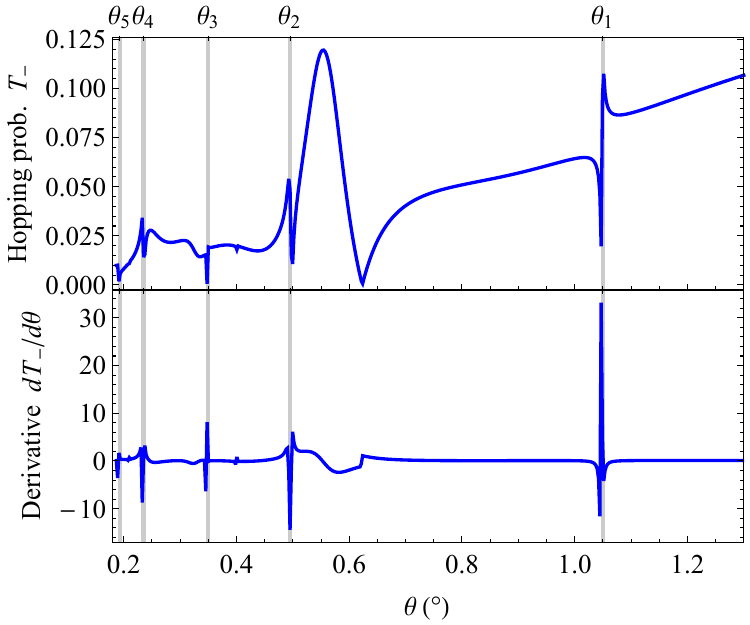}
	\end{center}
	\caption{Interlayer hopping probability (top panel) and its derivative (bottom panel)  for the lowest-energy valence (similar results for conduction) eigenstate $\psi_0^-$ as a function of the twist angle (in degrees). Magic angles are marked with vertical gray grid lines.  }
	\label{figProbHopping}
\end{figure}

\begin{figure}
	\begin{center}
\includegraphics[width=8cm]{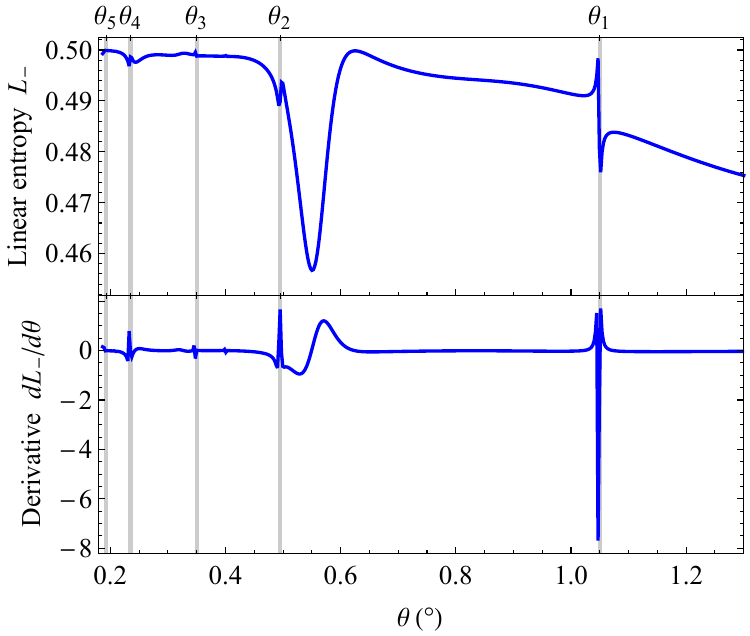}
	\end{center}
	\caption{Linear entropy  and its derivative for the lowest-energy valence (similar results for conduction) eigenstate $\psi_0^-$ as a function of the twist angle. Magic angles are marked with vertical gray grid lines.  }
	\label{figEntropy}
\end{figure}

Now we want to explore those quantities as a function of $\theta$ at the Dirac point $K_m$ ($\bm{q}=\bm{0}$). In Fig.~\ref{figProbHopping} we represent the interlayer hopping probability $T(\theta)$  and its derivative $dT/d\theta$   for the valence state $\psi_0^-$ (similar behavior for conduction  $\psi_0^+$) as a function of the twist angle $\theta$.   The transfer probability $T$ in general shows an oscillatory character for small $\theta$, but those oscillations are sharper (``resonances'')  at the magic angles, as can be seen by looking to the derivative $dT/d\theta$. Bernal stacking results at the Dirac point,  ${T}\to 0$ when $\theta\to 0$, are also recovered. A similar behavior is also shown in Fig. \ref{figEntropy} by the linear entropy $L(\theta)$, with sharper resonances  mainly at the principal magic angle $\theta_1=1.05^\circ$. Bernal stacking results at the Dirac point, $L\to 1/2$ when $\theta\to 0$, are also recovered. 

\begin{figure}
	\begin{center}
	\includegraphics[width=8cm]{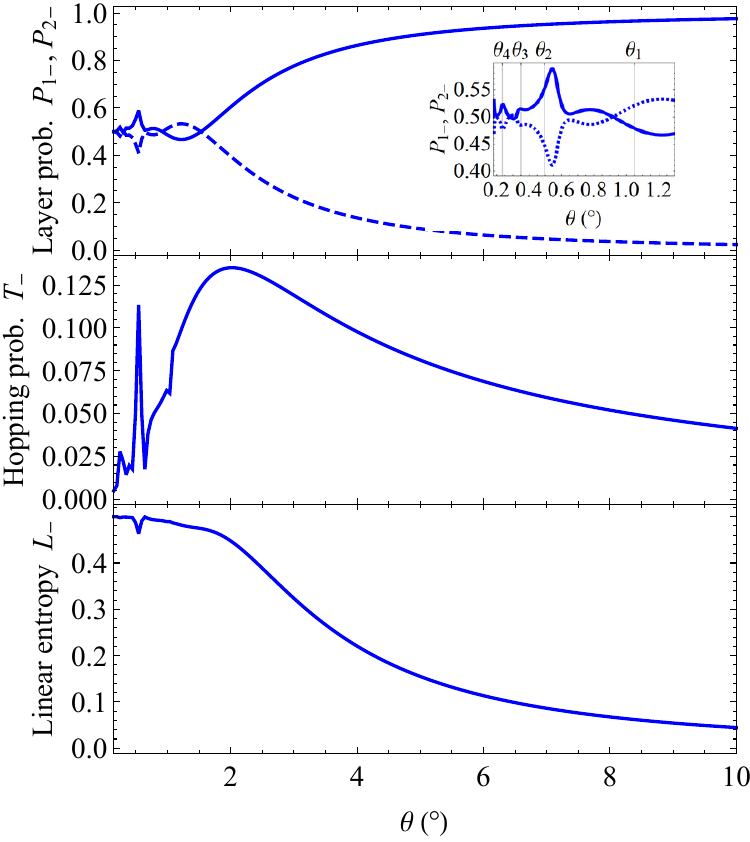}			
	\end{center}
	\caption{Large twist angle behavior of the layer probabilities $P_{\ell,-}$, the hopping probability $T_-$, and the linear entropy $L_-$, for the lowest-energy valence state $\psi_0^-$ (similar results are obtained for the conduction state). Inset in top figure shows layer  probabilities for the lowest-energy valence eigenstate $\psi_0^-$  (similar results for the conduction state $\psi_0^+$) as a function of the twist angle in the reduced interval $\theta\in[0.18,1.3]$ in degrees. Solid for layer $\ell=1$ and dashed for layer $\ell=2$.}
	\label{figLargeAngle}
\end{figure}

Layer probabilities $P_\ell$ in the inset of Fig. \ref{figLargeAngle} also show an oscillatory character with $\theta$ but they do not seem to capture magic angles. Bernal stacking results at the Dirac point, ${P}_1={P}_2\to 1/2$ when $\theta\to 0$, are also recovered. For completeness, and even though our Hamiltonian is valid for small enough twist angles $\theta$, in Fig. \ref{figLargeAngle} we show the large angle behavior of the RDM quantities $P_\ell, T$ and $L$. Oscillations and resonances seem to stop beyond the principal magic angle $\theta>\theta_1$. Extrapolating results beyond $\theta=10^\circ$, it seems  to happen that, for $\theta\gg 1$,  	
	${P}_1\to 1, {P}_2\to 0$, ${T}\to 0$ and 	
	$L\to0$. That is, the electron prefers one layer over the other and the lowest-energy Hamiltonian eigenstates $\psi_0^\pm$ are not entangled.

For completeness, and to support the idea that these results do not depend on the particular  entanglement measure used, we also provide the results for the Schlienz \& Mahler \cite{Mahler,mahlerBook} entanglement measure, as we did for Bernal stacking in previous section.  In this case, Hamiltonian eigenspinors have $D=2N$ components, with $D$ the Hamiltonian matrix size. We shall look at the TBG as a qubit-quNit system. The $2N\times 2N$ density matrix now reads
\begin{eqnarray}
\rho &=& \frac{1}{2N}\sigma_0 \otimes \Sigma_0+ \frac{1}{2N} \sum_{k=1}^3 \lambda^{(1)}_k \sigma_k \otimes \Sigma_0\\ && +   \frac{1}{4} \sum_{k=1}^{N^2-1}\lambda^{(2)}_k \sigma_0 \otimes \Sigma_k+  \frac{1}{4} \sum_{k= 1}^3\sum_{j=1}^{N^2-1} C^{(1,2)}_{kj} \sigma_k \otimes \Sigma_{j},\nn
\end{eqnarray}
where $\Sigma_j, j=1,\dots, N^2-1$ are generalized Gell-Mann matrices for SU($N$), that is, a basis of traceless hermitian $N\times N$ matrices fulfilling the orthogonality conditions $\tr(\Sigma_k\Sigma_l)=2\delta_{kl}$, and by  $\Sigma_0$ we denote the $N\times N$ identity matrix. The matrix $M$ in \eqref{MMahler} is now a $3\times (N^2-1)$ matrix and $\beta$ must be  normalized as
\begin{equation}
 \beta = \frac{1}{3} \tr({M}^T {M} ),
 \label{betaMahler}
\end{equation}
in order to have $0\leq\beta\leq 1$.

\begin{figure}
	\begin{center}
		\includegraphics[width=\columnwidth]{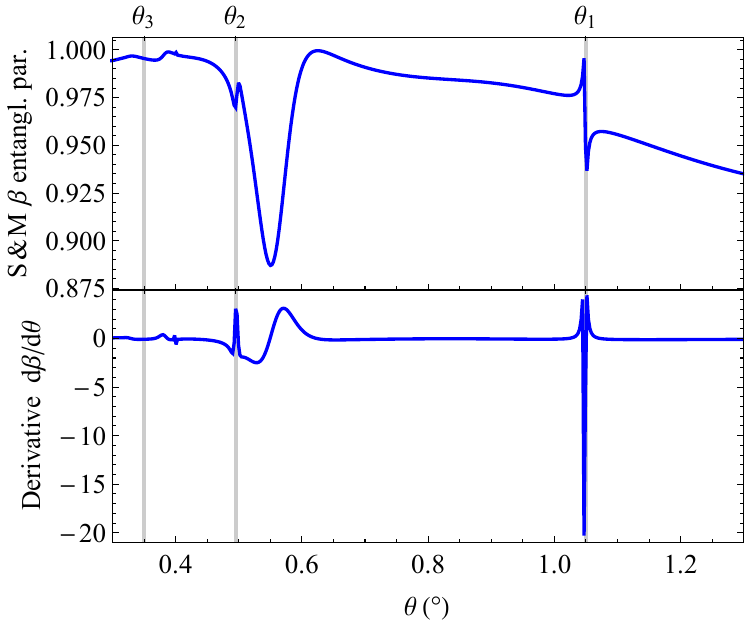}
	\end{center}
	\caption{Interlayer entanglement in TBG as a function of the twist angle $\theta$ measured in terms of Schlienz \& Mahler $\beta$ parameter \eqref{betaMahler} and its derivative. We use a Hamiltonian matrix size of $D=460$ (radius $R=10$) and the density matrix of the lowest-energy valence eigenstate $\rho=|\psi_0^-\rangle\langle\psi_0^-|$ (similar results for conduction). 
	}
	\label{figBeta}
\end{figure}

In Figure~\ref{figBeta} we show a plot of $\beta$ as a function of the twist angle $\theta$ taking as density matrix $\rho=|\psi_0^-\rangle\langle\psi_0^-|$ (similar results are obtained for the conduction state $|\psi_0^+$ at the Dirac point). For computational reasons, we have also restricted the calculations to radius $R=10$ and dimension $D=460$, so that we achieve converge up to  $\theta\gtrsim \sqrt{10}/R\sim 0.3^{\circ}$. The behavior of the $\beta$ parameter is similar to the interlayer hopping parameter and linear entropy in Figures \ref{figProbHopping},\ref{figEntropy}, that is, $\beta$ has sharp oscillations at the magic angles which are more evident in the derivative (bottom panel in Figure \ref{figBeta}). Therefore,  Schlienz \& Mahler $\beta$ parameter provides another entanglement measure detecting magic angles in TBG. In addition, the $\beta$ parameter approaches to 1 for small angles $\theta\to 0$, meeting the maximum entanglement limit in a Bernal-stacked form \eqref{BernalBeta} at the Dirac point $\bm{q}=0$.

\section{Conclusions}\label{conclusec}

We have contributed to unraveling the magic behind the electronic properties and exotic quantum states of TBG, which make it a promising platform for future device applications  in electronics,  optoelectronics, and quantum computing. Tools from the field of information theory offer a framework for examining and comprehending the complex interaction of degrees of freedom in TBG. Integrating these tools with quantum many-body physics and sophisticated computational techniques offers a promising strategy for investigating the rich landscape of quantum states in TBG and topological materials in general \cite{Jiang2012,ZengEntangTPT19}.

We have reviewed the flattening of low-energy Moir\'{e} Bloch bands and the formation of sharp picks in the DOS at magic angles $\theta\simeq 1.05^\circ, 0.495^\circ, 0.35^\circ, 0.235^\circ$ and $ 0.193^\circ$. This flattening is accompanied by a vanishing of the effective Dirac-point Fermi velocity for zero energy  Hamiltonian eigenstates $\psi_0(\theta)$ at those magic angles. Zero modes undergo critical structural changes at magic angles, and these abrupt changes are clearly detected by fidelity measurements. Fidelity also captures an anomalous behavior  at $\theta\simeq 0.4^\circ$, where the energies of the zero modes $\psi_0(\theta)$ and first excited $\psi_1(\theta)$ states degenerate, thus causing a spurious value of the Fermi velocity due to a numerical artifact.  Sharp oscillations (like ``resonances'') at magic angles  are also noticeable (specially at the principal one $\theta_1\simeq 1.05^\circ$) in the entanglement entropy and the  interlayer hopping probabilities (off-diagonal component of the RDM to the layer sector) as a function of the the twist angle. These resonances are better detected by the corresponding derivatives. Entanglement entropy is almost maximum in the small angle region where magic angles fall, and it decreases to zero for larger values of $\theta$. Low-energy  electrons seem to prefer one layer over the other for large $\theta$, as seen in the layer probabilities $P_\ell$, which oscillates around $P_\ell=0.5$ in the small angle region where magic angles fall. The interlayer hopping probability $T$ shows high variability for small angles and goes to zero for larger values of $\theta$.

High tunability with $\theta$ and manipulation with external electric and magnetic fields in general 2D materials allows for the exploration of novel quantum phenomena termed as ``twistronic'' \cite{PhysRevB.95.075420,Cao2018unconventional}. We believe that quantum information viewpoints can contribute to a better understanding for the physics behind and to the design of new experiments. 
Sudden transitions or abrupt changes in entanglement measurements, like those evidenced in this article at magic angles, can provide a wealth of opportunities for advancing quantum information devices and opens pathways to more robust, adaptable, and efficient quantum technologies. Monitoring such variations can provide a  diagnostic for fault-tolerant quantum error correction protocols. Also, sudden entanglement transitions often correspond to critical points in a quantum system (e.g., phase transitions); these critical points can be highly sensitive to external perturbations, making them ideal for quantum sensing applications and devices to achieve higher precision in measuring physical quantities like magnetic fields, gravitational waves, time intervals, etc.  Also to enhance the performance and reliability of quantum communication systems and  secure authentication in quantum cryptographic protocols. Bilayer graphene quantum dots  have proved  to reliably encode quantum information in the valley degree of freedom \cite{Garreis2024}, making it a promising quantum computation platform for electrically controllable robust qubits. Tunability with $\theta$ can offer here new possibilities since we have seen that interlayer hopping probability $T$ and entanglement entropy  show noticeable  resonances at magic angles (especially at $\theta_1\simeq 1.05^\circ$).

Novel quantum phenomena and new technological possibilities are also expected when adding a stacking degree of freedom and going from bilayer to multilayer arrangements. For example,  twisted trilayer graphene seems to provide a promising platform for experimental realization of correlated topological states  \cite{Park2021,HTG}. Rhombohedral pentalayer graphene provides an ideal platform for exploring charge fractionalization and (non-Abelian) anyonic braiding at zero magnetic field \cite{Lu2024}. Multilayer arrangements also increase the number of internal quantum degrees of freedom to the electron necessary to encode logical qubits. Therefore, investigations in this direction are worth to explore.

\emph{Note added: }  during the period  of review of this article, we have been informed of the existence of calculations based on atomistic tight-binding models  \cite{PhysRevResearch.1.013001,HungNguyen_2021}, including atomic relaxation  in the TBG model, where second (and higher) magic twist angles smaller than the principal one $\theta_1\simeq 1.1^{\circ}$ are not predicted.  Atomic reconstruction
is negligible for large twist angles but it turns out to be  essential for
small ones since it significantly modifies the stacking
structure of TBGs, compared to the ideal isolated
graphene lattices (see e.g. \cite{Xu2019,Yoo2019} for experiments). This fact deserves a separate study. Our main goal in this article has been to provide model-independent information-theoretic measures to capture all magic angles whenever they are present in a particular TBG Hamiltonian model.

\section*{Acknowledgments}
 We thank the support of Spanish MICIU through the project PID2022-138144NB-I00. AM thanks the Spanish MIU for the predoctoral grant FPU19/06376 at the University of Granada in the early stages of this work, and also thanks the Polytechnic University of Cartagena for a  contract under the European Commission project C-QuENS (ID: 101135359). OC was on sabbatical leave at Granada University, Spain,
from September 2023 to August 2024. OC thanks support from the program PASPA from DGAPA-UNAM.
 

\bibliography{bibliografia.bib}

\begin{thebibliography}{60}%
\makeatletter
\providecommand \@ifxundefined [1]{%
 \@ifx{#1\undefined}
}%
\providecommand \@ifnum [1]{%
 \ifnum #1\expandafter \@firstoftwo
 \else \expandafter \@secondoftwo
 \fi
}%
\providecommand \@ifx [1]{%
 \ifx #1\expandafter \@firstoftwo
 \else \expandafter \@secondoftwo
 \fi
}%
\providecommand \natexlab [1]{#1}%
\providecommand \enquote  [1]{``#1''}%
\providecommand \bibnamefont  [1]{#1}%
\providecommand \bibfnamefont [1]{#1}%
\providecommand \citenamefont [1]{#1}%
\providecommand \href@noop [0]{\@secondoftwo}%
\providecommand \href [0]{\begingroup \@sanitize@url \@href}%
\providecommand \@href[1]{\@@startlink{#1}\@@href}%
\providecommand \@@href[1]{\endgroup#1\@@endlink}%
\providecommand \@sanitize@url [0]{\catcode `\\12\catcode `\$12\catcode
  `\&12\catcode `\#12\catcode `\^12\catcode `\_12\catcode `\%12\relax}%
\providecommand \@@startlink[1]{}%
\providecommand \@@endlink[0]{}%
\providecommand \url  [0]{\begingroup\@sanitize@url \@url }%
\providecommand \@url [1]{\endgroup\@href {#1}{\urlprefix }}%
\providecommand \urlprefix  [0]{URL }%
\providecommand \Eprint [0]{\href }%
\providecommand \doibase [0]{https://doi.org/}%
\providecommand \selectlanguage [0]{\@gobble}%
\providecommand \bibinfo  [0]{\@secondoftwo}%
\providecommand \bibfield  [0]{\@secondoftwo}%
\providecommand \translation [1]{[#1]}%
\providecommand \BibitemOpen [0]{}%
\providecommand \bibitemStop [0]{}%
\providecommand \bibitemNoStop [0]{.\EOS\space}%
\providecommand \EOS [0]{\spacefactor3000\relax}%
\providecommand \BibitemShut  [1]{\csname bibitem#1\endcsname}%
\let\auto@bib@innerbib\@empty
\bibitem [{\citenamefont {Novoselov}\ \emph {et~al.}(2004)\citenamefont
  {Novoselov}, \citenamefont {Geim}, \citenamefont {Morozov}, \citenamefont
  {Jiang}, \citenamefont {Zhang}, \citenamefont {Dubonos}, \citenamefont
  {Grigorieva},\ and\ \citenamefont {Firsov}}]{Novoselov2004Science306}%
  \BibitemOpen
  \bibfield  {author} {\bibinfo {author} {\bibfnamefont {K.~S.}\ \bibnamefont
  {Novoselov}}, \bibinfo {author} {\bibfnamefont {A.~K.}\ \bibnamefont {Geim}},
  \bibinfo {author} {\bibfnamefont {S.~V.}\ \bibnamefont {Morozov}}, \bibinfo
  {author} {\bibfnamefont {D.}~\bibnamefont {Jiang}}, \bibinfo {author}
  {\bibfnamefont {Y.}~\bibnamefont {Zhang}}, \bibinfo {author} {\bibfnamefont
  {S.~V.}\ \bibnamefont {Dubonos}}, \bibinfo {author} {\bibfnamefont {I.~V.}\
  \bibnamefont {Grigorieva}},\ and\ \bibinfo {author} {\bibfnamefont {A.~A.}\
  \bibnamefont {Firsov}},\ }\bibfield  {title} {\bibinfo {title} {Electric
  field effect in atomically thin carbon films},\ }\href
  {https://doi.org/10.1126/science.1102896} {\bibfield  {journal} {\bibinfo
  {journal} {Science}\ }\textbf {\bibinfo {volume} {306}},\ \bibinfo {pages}
  {666} (\bibinfo {year} {2004})}\BibitemShut {NoStop}%
\bibitem [{\citenamefont {Novoselov}\ \emph {et~al.}(2005)\citenamefont
  {Novoselov}, \citenamefont {Jiang}, \citenamefont {Schedin}, \citenamefont
  {Booth}, \citenamefont {Khotkevich}, \citenamefont {Morozov},\ and\
  \citenamefont {Geim}}]{Novoselov2005PNAS102}%
  \BibitemOpen
  \bibfield  {author} {\bibinfo {author} {\bibfnamefont {K.~S.}\ \bibnamefont
  {Novoselov}}, \bibinfo {author} {\bibfnamefont {D.}~\bibnamefont {Jiang}},
  \bibinfo {author} {\bibfnamefont {F.}~\bibnamefont {Schedin}}, \bibinfo
  {author} {\bibfnamefont {T.~J.}\ \bibnamefont {Booth}}, \bibinfo {author}
  {\bibfnamefont {V.~V.}\ \bibnamefont {Khotkevich}}, \bibinfo {author}
  {\bibfnamefont {S.~V.}\ \bibnamefont {Morozov}},\ and\ \bibinfo {author}
  {\bibfnamefont {A.~K.}\ \bibnamefont {Geim}},\ }\bibfield  {title} {\bibinfo
  {title} {Two-dimensional atomic crystals},\ }\href
  {https://doi.org/10.1073/pnas.0502848102} {\bibfield  {journal} {\bibinfo
  {journal} {Proceedings of the National Academy of Sciences}\ }\textbf
  {\bibinfo {volume} {102}},\ \bibinfo {pages} {10451} (\bibinfo {year}
  {2005})}\BibitemShut {NoStop}%
\bibitem [{\citenamefont {Castro~Neto}\ \emph {et~al.}(2009)\citenamefont
  {Castro~Neto}, \citenamefont {Guinea}, \citenamefont {Peres}, \citenamefont
  {Novoselov},\ and\ \citenamefont {Geim}}]{RevModPhys.81.109}%
  \BibitemOpen
  \bibfield  {author} {\bibinfo {author} {\bibfnamefont {A.~H.}\ \bibnamefont
  {Castro~Neto}}, \bibinfo {author} {\bibfnamefont {F.}~\bibnamefont {Guinea}},
  \bibinfo {author} {\bibfnamefont {N.~M.~R.}\ \bibnamefont {Peres}}, \bibinfo
  {author} {\bibfnamefont {K.~S.}\ \bibnamefont {Novoselov}},\ and\ \bibinfo
  {author} {\bibfnamefont {A.~K.}\ \bibnamefont {Geim}},\ }\bibfield  {title}
  {\bibinfo {title} {The electronic properties of graphene},\ }\href
  {https://doi.org/10.1103/RevModPhys.81.109} {\bibfield  {journal} {\bibinfo
  {journal} {Rev. Mod. Phys.}\ }\textbf {\bibinfo {volume} {81}},\ \bibinfo
  {pages} {109} (\bibinfo {year} {2009})}\BibitemShut {NoStop}%
\bibitem [{\citenamefont {Kane}\ and\ \citenamefont {Mele}(2005)}]{KaneMele05}%
  \BibitemOpen
  \bibfield  {author} {\bibinfo {author} {\bibfnamefont {C.~L.}\ \bibnamefont
  {Kane}}\ and\ \bibinfo {author} {\bibfnamefont {E.~J.}\ \bibnamefont
  {Mele}},\ }\bibfield  {title} {\bibinfo {title} {Quantum spin hall effect in
  graphene},\ }\href {https://doi.org/10.1103/PhysRevLett.95.226801} {\bibfield
   {journal} {\bibinfo  {journal} {Phys. Rev. Lett.}\ }\textbf {\bibinfo
  {volume} {95}},\ \bibinfo {pages} {226801} (\bibinfo {year}
  {2005})}\BibitemShut {NoStop}%
\bibitem [{\citenamefont {Hasan}\ and\ \citenamefont
  {Kane}(2010)}]{RevModPhys.82.3045}%
  \BibitemOpen
  \bibfield  {author} {\bibinfo {author} {\bibfnamefont {M.~Z.}\ \bibnamefont
  {Hasan}}\ and\ \bibinfo {author} {\bibfnamefont {C.~L.}\ \bibnamefont
  {Kane}},\ }\bibfield  {title} {\bibinfo {title} {Colloquium: Topological
  insulators},\ }\href {https://doi.org/10.1103/RevModPhys.82.3045} {\bibfield
  {journal} {\bibinfo  {journal} {Rev. Mod. Phys.}\ }\textbf {\bibinfo {volume}
  {82}},\ \bibinfo {pages} {3045} (\bibinfo {year} {2010})}\BibitemShut
  {NoStop}%
\bibitem [{\citenamefont {Qi}\ and\ \citenamefont
  {Zhang}(2011)}]{RevModPhys.83.1057}%
  \BibitemOpen
  \bibfield  {author} {\bibinfo {author} {\bibfnamefont {X.-L.}\ \bibnamefont
  {Qi}}\ and\ \bibinfo {author} {\bibfnamefont {S.-C.}\ \bibnamefont {Zhang}},\
  }\bibfield  {title} {\bibinfo {title} {Topological insulators and
  superconductors},\ }\href {https://doi.org/10.1103/RevModPhys.83.1057}
  {\bibfield  {journal} {\bibinfo  {journal} {Rev. Mod. Phys.}\ }\textbf
  {\bibinfo {volume} {83}},\ \bibinfo {pages} {1057–1110} (\bibinfo {year}
  {2011})}\BibitemShut {NoStop}%
\bibitem [{\citenamefont {Icking}\ \emph {et~al.}(2024)\citenamefont {Icking},
  \citenamefont {Emmerich}, \citenamefont {Watanabe}, \citenamefont
  {Taniguchi}, \citenamefont {Beschoten}, \citenamefont {Lemme}, \citenamefont
  {Knoch},\ and\ \citenamefont {Stampfer}}]{Icking2024}%
  \BibitemOpen
  \bibfield  {author} {\bibinfo {author} {\bibfnamefont {E.}~\bibnamefont
  {Icking}}, \bibinfo {author} {\bibfnamefont {D.}~\bibnamefont {Emmerich}},
  \bibinfo {author} {\bibfnamefont {K.}~\bibnamefont {Watanabe}}, \bibinfo
  {author} {\bibfnamefont {T.}~\bibnamefont {Taniguchi}}, \bibinfo {author}
  {\bibfnamefont {B.}~\bibnamefont {Beschoten}}, \bibinfo {author}
  {\bibfnamefont {M.~C.}\ \bibnamefont {Lemme}}, \bibinfo {author}
  {\bibfnamefont {J.}~\bibnamefont {Knoch}},\ and\ \bibinfo {author}
  {\bibfnamefont {C.}~\bibnamefont {Stampfer}},\ }\bibfield  {title} {\bibinfo
  {title} {Ultrasteep slope cryogenic fets based on bilayer graphene},\
  }\bibfield  {booktitle} {\emph {\bibinfo {booktitle} {Nano Letters}},\ }\href
  {https://doi.org/10.1021/acs.nanolett.4c02463} {\bibfield  {journal}
  {\bibinfo  {journal} {Nano Letters}\ }\textbf {\bibinfo {volume} {24}},\
  \bibinfo {pages} {11454} (\bibinfo {year} {2024})}\BibitemShut {NoStop}%
\bibitem [{\citenamefont {Rozhkov}\ \emph {et~al.}(2016)\citenamefont
  {Rozhkov}, \citenamefont {Sboychakov}, \citenamefont {Rakhmanov},\ and\
  \citenamefont {Nori}}]{ROZHKOV20161}%
  \BibitemOpen
  \bibfield  {author} {\bibinfo {author} {\bibfnamefont {A.}~\bibnamefont
  {Rozhkov}}, \bibinfo {author} {\bibfnamefont {A.}~\bibnamefont {Sboychakov}},
  \bibinfo {author} {\bibfnamefont {A.}~\bibnamefont {Rakhmanov}},\ and\
  \bibinfo {author} {\bibfnamefont {F.}~\bibnamefont {Nori}},\ }\bibfield
  {title} {\bibinfo {title} {Electronic properties of graphene-based bilayer
  systems},\ }\href
  {https://doi.org/https://doi.org/10.1016/j.physrep.2016.07.003} {\bibfield
  {journal} {\bibinfo  {journal} {Physics Reports}\ }\textbf {\bibinfo {volume}
  {648}},\ \bibinfo {pages} {1} (\bibinfo {year} {2016})}\BibitemShut {NoStop}%
\bibitem [{\citenamefont {Cao}\ \emph {et~al.}(2018{\natexlab{a}})\citenamefont
  {Cao}, \citenamefont {Fatemi}, \citenamefont {Fang}, \citenamefont
  {Watanabe}, \citenamefont {Taniguchi}, \citenamefont {Kaxiras},\ and\
  \citenamefont {Jarillo-Herrero}}]{Cao2018unconventional}%
  \BibitemOpen
  \bibfield  {author} {\bibinfo {author} {\bibfnamefont {Y.}~\bibnamefont
  {Cao}}, \bibinfo {author} {\bibfnamefont {V.}~\bibnamefont {Fatemi}},
  \bibinfo {author} {\bibfnamefont {S.}~\bibnamefont {Fang}}, \bibinfo {author}
  {\bibfnamefont {K.}~\bibnamefont {Watanabe}}, \bibinfo {author}
  {\bibfnamefont {T.}~\bibnamefont {Taniguchi}}, \bibinfo {author}
  {\bibfnamefont {E.}~\bibnamefont {Kaxiras}},\ and\ \bibinfo {author}
  {\bibfnamefont {P.}~\bibnamefont {Jarillo-Herrero}},\ }\bibfield  {title}
  {\bibinfo {title} {Unconventional superconductivity in magic-angle graphene
  superlattices},\ }\href {https://doi.org/10.1038/nature26160} {\bibfield
  {journal} {\bibinfo  {journal} {Nature}\ }\textbf {\bibinfo {volume} {556}},\
  \bibinfo {pages} {43} (\bibinfo {year} {2018}{\natexlab{a}})}\BibitemShut
  {NoStop}%
\bibitem [{\citenamefont {Cao}\ \emph {et~al.}(2018{\natexlab{b}})\citenamefont
  {Cao}, \citenamefont {Fatemi}, \citenamefont {Demir}, \citenamefont {Fang},
  \citenamefont {Tomarken}, \citenamefont {Luo}, \citenamefont
  {Sanchez-Yamagishi}, \citenamefont {Watanabe}, \citenamefont {Taniguchi},
  \citenamefont {Kaxiras}, \citenamefont {Ashoori},\ and\ \citenamefont
  {Jarillo-Herrero}}]{Cao2018correlated}%
  \BibitemOpen
  \bibfield  {author} {\bibinfo {author} {\bibfnamefont {Y.}~\bibnamefont
  {Cao}}, \bibinfo {author} {\bibfnamefont {V.}~\bibnamefont {Fatemi}},
  \bibinfo {author} {\bibfnamefont {A.}~\bibnamefont {Demir}}, \bibinfo
  {author} {\bibfnamefont {S.}~\bibnamefont {Fang}}, \bibinfo {author}
  {\bibfnamefont {S.~L.}\ \bibnamefont {Tomarken}}, \bibinfo {author}
  {\bibfnamefont {J.~Y.}\ \bibnamefont {Luo}}, \bibinfo {author} {\bibfnamefont
  {J.~D.}\ \bibnamefont {Sanchez-Yamagishi}}, \bibinfo {author} {\bibfnamefont
  {K.}~\bibnamefont {Watanabe}}, \bibinfo {author} {\bibfnamefont
  {T.}~\bibnamefont {Taniguchi}}, \bibinfo {author} {\bibfnamefont
  {E.}~\bibnamefont {Kaxiras}}, \bibinfo {author} {\bibfnamefont {R.~C.}\
  \bibnamefont {Ashoori}},\ and\ \bibinfo {author} {\bibfnamefont
  {P.}~\bibnamefont {Jarillo-Herrero}},\ }\bibfield  {title} {\bibinfo {title}
  {Correlated insulator behaviour at half-filling in magic-angle graphene
  superlattices},\ }\href {https://doi.org/10.1038/nature26154} {\bibfield
  {journal} {\bibinfo  {journal} {Nature}\ }\textbf {\bibinfo {volume} {556}},\
  \bibinfo {pages} {80} (\bibinfo {year} {2018}{\natexlab{b}})}\BibitemShut
  {NoStop}%
\bibitem [{\citenamefont {Bistritzer}\ and\ \citenamefont
  {MacDonald}(2011)}]{Bistritzer12233}%
  \BibitemOpen
  \bibfield  {author} {\bibinfo {author} {\bibfnamefont {R.}~\bibnamefont
  {Bistritzer}}\ and\ \bibinfo {author} {\bibfnamefont {A.~H.}\ \bibnamefont
  {MacDonald}},\ }\bibfield  {title} {\bibinfo {title} {Moiré bands in twisted
  double-layer graphene},\ }\href {https://doi.org/10.1073/pnas.1108174108}
  {\bibfield  {journal} {\bibinfo  {journal} {Proceedings of the National
  Academy of Sciences}\ }\textbf {\bibinfo {volume} {108}},\ \bibinfo {pages}
  {12233–12237} (\bibinfo {year} {2011})}\BibitemShut {NoStop}%
\bibitem [{\citenamefont {Tabert}\ and\ \citenamefont
  {Nicol}(2013)}]{PhysRevB.87.121402}%
  \BibitemOpen
  \bibfield  {author} {\bibinfo {author} {\bibfnamefont {C.~J.}\ \bibnamefont
  {Tabert}}\ and\ \bibinfo {author} {\bibfnamefont {E.~J.}\ \bibnamefont
  {Nicol}},\ }\bibfield  {title} {\bibinfo {title} {Optical conductivity of
  twisted bilayer graphene},\ }\href
  {https://doi.org/10.1103/PhysRevB.87.121402} {\bibfield  {journal} {\bibinfo
  {journal} {Phys. Rev. B}\ }\textbf {\bibinfo {volume} {87}},\ \bibinfo
  {pages} {121402(R)} (\bibinfo {year} {2013})}\BibitemShut {NoStop}%
\bibitem [{\citenamefont {Li}\ \emph {et~al.}(2010)\citenamefont {Li},
  \citenamefont {Luican}, \citenamefont {Lopes~dos Santos}, \citenamefont
  {Castro~Neto}, \citenamefont {Reina}, \citenamefont {Kong},\ and\
  \citenamefont {Andrei}}]{Li2010}%
  \BibitemOpen
  \bibfield  {author} {\bibinfo {author} {\bibfnamefont {G.}~\bibnamefont
  {Li}}, \bibinfo {author} {\bibfnamefont {A.}~\bibnamefont {Luican}}, \bibinfo
  {author} {\bibfnamefont {J.~M.~B.}\ \bibnamefont {Lopes~dos Santos}},
  \bibinfo {author} {\bibfnamefont {A.~H.}\ \bibnamefont {Castro~Neto}},
  \bibinfo {author} {\bibfnamefont {A.}~\bibnamefont {Reina}}, \bibinfo
  {author} {\bibfnamefont {J.}~\bibnamefont {Kong}},\ and\ \bibinfo {author}
  {\bibfnamefont {E.~Y.}\ \bibnamefont {Andrei}},\ }\bibfield  {title}
  {\bibinfo {title} {Observation of van hove singularities in twisted graphene
  layers},\ }\href {https://doi.org/10.1038/nphys1463} {\bibfield  {journal}
  {\bibinfo  {journal} {Nature Physics}\ }\textbf {\bibinfo {volume} {6}},\
  \bibinfo {pages} {109} (\bibinfo {year} {2010})}\BibitemShut {NoStop}%
\bibitem [{\citenamefont {Brihuega}\ \emph {et~al.}(2012)\citenamefont
  {Brihuega}, \citenamefont {Mallet}, \citenamefont {Gonz\'alez-Herrero},
  \citenamefont {Trambly~de Laissardi\`ere}, \citenamefont {Ugeda},
  \citenamefont {Magaud}, \citenamefont {G\'omez-Rodr\'{\i}guez}, \citenamefont
  {Yndur\'ain},\ and\ \citenamefont {Veuillen}}]{PhysRevLett.109.196802}%
  \BibitemOpen
  \bibfield  {author} {\bibinfo {author} {\bibfnamefont {I.}~\bibnamefont
  {Brihuega}}, \bibinfo {author} {\bibfnamefont {P.}~\bibnamefont {Mallet}},
  \bibinfo {author} {\bibfnamefont {H.}~\bibnamefont {Gonz\'alez-Herrero}},
  \bibinfo {author} {\bibfnamefont {G.}~\bibnamefont {Trambly~de
  Laissardi\`ere}}, \bibinfo {author} {\bibfnamefont {M.~M.}\ \bibnamefont
  {Ugeda}}, \bibinfo {author} {\bibfnamefont {L.}~\bibnamefont {Magaud}},
  \bibinfo {author} {\bibfnamefont {J.~M.}\ \bibnamefont
  {G\'omez-Rodr\'{\i}guez}}, \bibinfo {author} {\bibfnamefont {F.}~\bibnamefont
  {Yndur\'ain}},\ and\ \bibinfo {author} {\bibfnamefont {J.-Y.}\ \bibnamefont
  {Veuillen}},\ }\bibfield  {title} {\bibinfo {title} {Unraveling the intrinsic
  and robust nature of van hove singularities in twisted bilayer graphene by
  scanning tunneling microscopy and theoretical analysis},\ }\href
  {https://doi.org/10.1103/PhysRevLett.109.196802} {\bibfield  {journal}
  {\bibinfo  {journal} {Phys. Rev. Lett.}\ }\textbf {\bibinfo {volume} {109}},\
  \bibinfo {pages} {196802} (\bibinfo {year} {2012})}\BibitemShut {NoStop}%
\bibitem [{\citenamefont {Weckbecker}\ \emph {et~al.}(2016)\citenamefont
  {Weckbecker}, \citenamefont {Shallcross}, \citenamefont {Fleischmann},
  \citenamefont {Ray}, \citenamefont {Sharma},\ and\ \citenamefont
  {Pankratov}}]{PhysRevB.93.035452}%
  \BibitemOpen
  \bibfield  {author} {\bibinfo {author} {\bibfnamefont {D.}~\bibnamefont
  {Weckbecker}}, \bibinfo {author} {\bibfnamefont {S.}~\bibnamefont
  {Shallcross}}, \bibinfo {author} {\bibfnamefont {M.}~\bibnamefont
  {Fleischmann}}, \bibinfo {author} {\bibfnamefont {N.}~\bibnamefont {Ray}},
  \bibinfo {author} {\bibfnamefont {S.}~\bibnamefont {Sharma}},\ and\ \bibinfo
  {author} {\bibfnamefont {O.}~\bibnamefont {Pankratov}},\ }\bibfield  {title}
  {\bibinfo {title} {Low-energy theory for the graphene twist bilayer},\ }\href
  {https://doi.org/10.1103/PhysRevB.93.035452} {\bibfield  {journal} {\bibinfo
  {journal} {Phys. Rev. B}\ }\textbf {\bibinfo {volume} {93}},\ \bibinfo
  {pages} {035452} (\bibinfo {year} {2016})}\BibitemShut {NoStop}%
\bibitem [{\citenamefont {de~Gail}\ \emph {et~al.}(2011)\citenamefont
  {de~Gail}, \citenamefont {Goerbig}, \citenamefont {Guinea}, \citenamefont
  {Montambaux},\ and\ \citenamefont {Castro~Neto}}]{PhysRevB.84.045436}%
  \BibitemOpen
  \bibfield  {author} {\bibinfo {author} {\bibfnamefont {R.}~\bibnamefont
  {de~Gail}}, \bibinfo {author} {\bibfnamefont {M.~O.}\ \bibnamefont
  {Goerbig}}, \bibinfo {author} {\bibfnamefont {F.}~\bibnamefont {Guinea}},
  \bibinfo {author} {\bibfnamefont {G.}~\bibnamefont {Montambaux}},\ and\
  \bibinfo {author} {\bibfnamefont {A.~H.}\ \bibnamefont {Castro~Neto}},\
  }\bibfield  {title} {\bibinfo {title} {Topologically protected zero modes in
  twisted bilayer graphene},\ }\href
  {https://doi.org/10.1103/PhysRevB.84.045436} {\bibfield  {journal} {\bibinfo
  {journal} {Phys. Rev. B}\ }\textbf {\bibinfo {volume} {84}},\ \bibinfo
  {pages} {045436} (\bibinfo {year} {2011})}\BibitemShut {NoStop}%
\bibitem [{\citenamefont {Song}\ \emph {et~al.}(2019)\citenamefont {Song},
  \citenamefont {Wang}, \citenamefont {Shi}, \citenamefont {Li}, \citenamefont
  {Fang},\ and\ \citenamefont {Bernevig}}]{PhysRevLett.123.036401}%
  \BibitemOpen
  \bibfield  {author} {\bibinfo {author} {\bibfnamefont {Z.}~\bibnamefont
  {Song}}, \bibinfo {author} {\bibfnamefont {Z.}~\bibnamefont {Wang}}, \bibinfo
  {author} {\bibfnamefont {W.}~\bibnamefont {Shi}}, \bibinfo {author}
  {\bibfnamefont {G.}~\bibnamefont {Li}}, \bibinfo {author} {\bibfnamefont
  {C.}~\bibnamefont {Fang}},\ and\ \bibinfo {author} {\bibfnamefont {B.~A.}\
  \bibnamefont {Bernevig}},\ }\bibfield  {title} {\bibinfo {title} {All magic
  angles in twisted bilayer graphene are topological},\ }\href
  {https://doi.org/10.1103/PhysRevLett.123.036401} {\bibfield  {journal}
  {\bibinfo  {journal} {Phys. Rev. Lett.}\ }\textbf {\bibinfo {volume} {123}},\
  \bibinfo {pages} {036401} (\bibinfo {year} {2019})}\BibitemShut {NoStop}%
\bibitem [{\citenamefont {Bernevig}\ \emph
  {et~al.}(2021{\natexlab{a}})\citenamefont {Bernevig}, \citenamefont {Song},
  \citenamefont {Regnault},\ and\ \citenamefont {Lian}}]{PhysRevB.103.205411}%
  \BibitemOpen
  \bibfield  {author} {\bibinfo {author} {\bibfnamefont {B.~A.}\ \bibnamefont
  {Bernevig}}, \bibinfo {author} {\bibfnamefont {Z.-D.}\ \bibnamefont {Song}},
  \bibinfo {author} {\bibfnamefont {N.}~\bibnamefont {Regnault}},\ and\
  \bibinfo {author} {\bibfnamefont {B.}~\bibnamefont {Lian}},\ }\bibfield
  {title} {\bibinfo {title} {Twisted bilayer graphene. i. matrix elements,
  approximations, perturbation theory, and a
  $k\ifmmode\cdot\else\textperiodcentered\fi{}p$ two-band model},\ }\href
  {https://doi.org/10.1103/PhysRevB.103.205411} {\bibfield  {journal} {\bibinfo
   {journal} {Phys. Rev. B}\ }\textbf {\bibinfo {volume} {103}},\ \bibinfo
  {pages} {205411} (\bibinfo {year} {2021}{\natexlab{a}})}\BibitemShut
  {NoStop}%
\bibitem [{\citenamefont {Song}\ \emph {et~al.}(2021)\citenamefont {Song},
  \citenamefont {Lian}, \citenamefont {Regnault},\ and\ \citenamefont
  {Bernevig}}]{PhysRevB.103.205412}%
  \BibitemOpen
  \bibfield  {author} {\bibinfo {author} {\bibfnamefont {Z.-D.}\ \bibnamefont
  {Song}}, \bibinfo {author} {\bibfnamefont {B.}~\bibnamefont {Lian}}, \bibinfo
  {author} {\bibfnamefont {N.}~\bibnamefont {Regnault}},\ and\ \bibinfo
  {author} {\bibfnamefont {B.~A.}\ \bibnamefont {Bernevig}},\ }\bibfield
  {title} {\bibinfo {title} {Twisted bilayer graphene. ii. stable symmetry
  anomaly},\ }\href {https://doi.org/10.1103/PhysRevB.103.205412} {\bibfield
  {journal} {\bibinfo  {journal} {Phys. Rev. B}\ }\textbf {\bibinfo {volume}
  {103}},\ \bibinfo {pages} {205412} (\bibinfo {year} {2021})}\BibitemShut
  {NoStop}%
\bibitem [{\citenamefont {Bernevig}\ \emph
  {et~al.}(2021{\natexlab{b}})\citenamefont {Bernevig}, \citenamefont {Song},
  \citenamefont {Regnault},\ and\ \citenamefont {Lian}}]{PhysRevB.103.205413}%
  \BibitemOpen
  \bibfield  {author} {\bibinfo {author} {\bibfnamefont {B.~A.}\ \bibnamefont
  {Bernevig}}, \bibinfo {author} {\bibfnamefont {Z.-D.}\ \bibnamefont {Song}},
  \bibinfo {author} {\bibfnamefont {N.}~\bibnamefont {Regnault}},\ and\
  \bibinfo {author} {\bibfnamefont {B.}~\bibnamefont {Lian}},\ }\bibfield
  {title} {\bibinfo {title} {Twisted bilayer graphene. iii. interacting
  hamiltonian and exact symmetries},\ }\href
  {https://doi.org/10.1103/PhysRevB.103.205413} {\bibfield  {journal} {\bibinfo
   {journal} {Phys. Rev. B}\ }\textbf {\bibinfo {volume} {103}},\ \bibinfo
  {pages} {205413} (\bibinfo {year} {2021}{\natexlab{b}})}\BibitemShut
  {NoStop}%
\bibitem [{\citenamefont {Lian}\ \emph {et~al.}(2021)\citenamefont {Lian},
  \citenamefont {Song}, \citenamefont {Regnault}, \citenamefont {Efetov},
  \citenamefont {Yazdani},\ and\ \citenamefont
  {Bernevig}}]{PhysRevB.103.205414}%
  \BibitemOpen
  \bibfield  {author} {\bibinfo {author} {\bibfnamefont {B.}~\bibnamefont
  {Lian}}, \bibinfo {author} {\bibfnamefont {Z.-D.}\ \bibnamefont {Song}},
  \bibinfo {author} {\bibfnamefont {N.}~\bibnamefont {Regnault}}, \bibinfo
  {author} {\bibfnamefont {D.~K.}\ \bibnamefont {Efetov}}, \bibinfo {author}
  {\bibfnamefont {A.}~\bibnamefont {Yazdani}},\ and\ \bibinfo {author}
  {\bibfnamefont {B.~A.}\ \bibnamefont {Bernevig}},\ }\bibfield  {title}
  {\bibinfo {title} {Twisted bilayer graphene. iv. exact insulator ground
  states and phase diagram},\ }\href
  {https://doi.org/10.1103/PhysRevB.103.205414} {\bibfield  {journal} {\bibinfo
   {journal} {Phys. Rev. B}\ }\textbf {\bibinfo {volume} {103}},\ \bibinfo
  {pages} {205414} (\bibinfo {year} {2021})}\BibitemShut {NoStop}%
\bibitem [{\citenamefont {Bernevig}\ \emph
  {et~al.}(2021{\natexlab{c}})\citenamefont {Bernevig}, \citenamefont {Lian},
  \citenamefont {Cowsik}, \citenamefont {Xie}, \citenamefont {Regnault},\ and\
  \citenamefont {Song}}]{PhysRevB.103.205415}%
  \BibitemOpen
  \bibfield  {author} {\bibinfo {author} {\bibfnamefont {B.~A.}\ \bibnamefont
  {Bernevig}}, \bibinfo {author} {\bibfnamefont {B.}~\bibnamefont {Lian}},
  \bibinfo {author} {\bibfnamefont {A.}~\bibnamefont {Cowsik}}, \bibinfo
  {author} {\bibfnamefont {F.}~\bibnamefont {Xie}}, \bibinfo {author}
  {\bibfnamefont {N.}~\bibnamefont {Regnault}},\ and\ \bibinfo {author}
  {\bibfnamefont {Z.-D.}\ \bibnamefont {Song}},\ }\bibfield  {title} {\bibinfo
  {title} {Twisted bilayer graphene. v. exact analytic many-body excitations in
  coulomb hamiltonians: Charge gap, goldstone modes, and absence of cooper
  pairing},\ }\href {https://doi.org/10.1103/PhysRevB.103.205415} {\bibfield
  {journal} {\bibinfo  {journal} {Phys. Rev. B}\ }\textbf {\bibinfo {volume}
  {103}},\ \bibinfo {pages} {205415} (\bibinfo {year}
  {2021}{\natexlab{c}})}\BibitemShut {NoStop}%
\bibitem [{\citenamefont {Xie}\ \emph {et~al.}(2021)\citenamefont {Xie},
  \citenamefont {Cowsik}, \citenamefont {Song}, \citenamefont {Lian},
  \citenamefont {Bernevig},\ and\ \citenamefont
  {Regnault}}]{PhysRevB.103.205416}%
  \BibitemOpen
  \bibfield  {author} {\bibinfo {author} {\bibfnamefont {F.}~\bibnamefont
  {Xie}}, \bibinfo {author} {\bibfnamefont {A.}~\bibnamefont {Cowsik}},
  \bibinfo {author} {\bibfnamefont {Z.-D.}\ \bibnamefont {Song}}, \bibinfo
  {author} {\bibfnamefont {B.}~\bibnamefont {Lian}}, \bibinfo {author}
  {\bibfnamefont {B.~A.}\ \bibnamefont {Bernevig}},\ and\ \bibinfo {author}
  {\bibfnamefont {N.}~\bibnamefont {Regnault}},\ }\bibfield  {title} {\bibinfo
  {title} {Twisted bilayer graphene. vi. an exact diagonalization study at
  nonzero integer filling},\ }\href
  {https://doi.org/10.1103/PhysRevB.103.205416} {\bibfield  {journal} {\bibinfo
   {journal} {Phys. Rev. B}\ }\textbf {\bibinfo {volume} {103}},\ \bibinfo
  {pages} {205416} (\bibinfo {year} {2021})}\BibitemShut {NoStop}%
\bibitem [{\citenamefont {Lopes~dos Santos}\ \emph {et~al.}(2007)\citenamefont
  {Lopes~dos Santos}, \citenamefont {Peres},\ and\ \citenamefont
  {Castro~Neto}}]{PhysRevLett.99.256802}%
  \BibitemOpen
  \bibfield  {author} {\bibinfo {author} {\bibfnamefont {J.~M.~B.}\
  \bibnamefont {Lopes~dos Santos}}, \bibinfo {author} {\bibfnamefont
  {N.~M.~R.}\ \bibnamefont {Peres}},\ and\ \bibinfo {author} {\bibfnamefont
  {A.~H.}\ \bibnamefont {Castro~Neto}},\ }\bibfield  {title} {\bibinfo {title}
  {Graphene bilayer with a twist: Electronic structure},\ }\href
  {https://doi.org/10.1103/PhysRevLett.99.256802} {\bibfield  {journal}
  {\bibinfo  {journal} {Phys. Rev. Lett.}\ }\textbf {\bibinfo {volume} {99}},\
  \bibinfo {pages} {256802} (\bibinfo {year} {2007})}\BibitemShut {NoStop}%
\bibitem [{\citenamefont {Lopes~dos Santos}\ \emph {et~al.}(2012)\citenamefont
  {Lopes~dos Santos}, \citenamefont {Peres},\ and\ \citenamefont
  {Castro~Neto}}]{PhysRevB.86.155449}%
  \BibitemOpen
  \bibfield  {author} {\bibinfo {author} {\bibfnamefont {J.~M.~B.}\
  \bibnamefont {Lopes~dos Santos}}, \bibinfo {author} {\bibfnamefont
  {N.~M.~R.}\ \bibnamefont {Peres}},\ and\ \bibinfo {author} {\bibfnamefont
  {A.~H.}\ \bibnamefont {Castro~Neto}},\ }\bibfield  {title} {\bibinfo {title}
  {Continuum model of the twisted graphene bilayer},\ }\href
  {https://doi.org/10.1103/PhysRevB.86.155449} {\bibfield  {journal} {\bibinfo
  {journal} {Phys. Rev. B}\ }\textbf {\bibinfo {volume} {86}},\ \bibinfo
  {pages} {155449} (\bibinfo {year} {2012})}\BibitemShut {NoStop}%
\bibitem [{\citenamefont {Tarnopolsky}\ \emph {et~al.}(2019)\citenamefont
  {Tarnopolsky}, \citenamefont {Kruchkov},\ and\ \citenamefont
  {Vishwanath}}]{PhysRevLett.122.106405}%
  \BibitemOpen
  \bibfield  {author} {\bibinfo {author} {\bibfnamefont {G.}~\bibnamefont
  {Tarnopolsky}}, \bibinfo {author} {\bibfnamefont {A.~J.}\ \bibnamefont
  {Kruchkov}},\ and\ \bibinfo {author} {\bibfnamefont {A.}~\bibnamefont
  {Vishwanath}},\ }\bibfield  {title} {\bibinfo {title} {Origin of magic angles
  in twisted bilayer graphene},\ }\href
  {https://doi.org/10.1103/PhysRevLett.122.106405} {\bibfield  {journal}
  {\bibinfo  {journal} {Phys. Rev. Lett.}\ }\textbf {\bibinfo {volume} {122}},\
  \bibinfo {pages} {106405} (\bibinfo {year} {2019})}\BibitemShut {NoStop}%
\bibitem [{\citenamefont {Jiang}\ \emph {et~al.}(2012)\citenamefont {Jiang},
  \citenamefont {Wang},\ and\ \citenamefont {Balents}}]{Jiang2012}%
  \BibitemOpen
  \bibfield  {author} {\bibinfo {author} {\bibfnamefont {H.-C.}\ \bibnamefont
  {Jiang}}, \bibinfo {author} {\bibfnamefont {Z.}~\bibnamefont {Wang}},\ and\
  \bibinfo {author} {\bibfnamefont {L.}~\bibnamefont {Balents}},\ }\bibfield
  {title} {\bibinfo {title} {Identifying topological order by entanglement
  entropy},\ }\href {https://doi.org/10.1038/nphys2465} {\bibfield  {journal}
  {\bibinfo  {journal} {Nature Physics}\ }\textbf {\bibinfo {volume} {8}},\
  \bibinfo {pages} {902–905} (\bibinfo {year} {2012})}\BibitemShut {NoStop}%
\bibitem [{\citenamefont {Zeng}\ \emph {et~al.}(2019)\citenamefont {Zeng},
  \citenamefont {Chen}, \citenamefont {Zhou},\ and\ \citenamefont
  {Wen}}]{ZengEntangTPT19}%
  \BibitemOpen
  \bibfield  {author} {\bibinfo {author} {\bibfnamefont {B.}~\bibnamefont
  {Zeng}}, \bibinfo {author} {\bibfnamefont {X.}~\bibnamefont {Chen}}, \bibinfo
  {author} {\bibfnamefont {D.-L.}\ \bibnamefont {Zhou}},\ and\ \bibinfo
  {author} {\bibfnamefont {X.-G.}\ \bibnamefont {Wen}},\ }\href
  {https://doi.org/10.1007/978-1-4939-9084-9} {\emph {\bibinfo {title} {Quantum
  Information Meets Quantum Matter: From Quantum Entanglement to Topological
  Phases of Many-Body Systems}}}\ (\bibinfo  {publisher} {Springer Nature},\
  \bibinfo {year} {2019})\BibitemShut {NoStop}%
\bibitem [{\citenamefont {Abanin}\ and\ \citenamefont
  {Demler}(2012)}]{PhysRevLett.109.020504}%
  \BibitemOpen
  \bibfield  {author} {\bibinfo {author} {\bibfnamefont {D.~A.}\ \bibnamefont
  {Abanin}}\ and\ \bibinfo {author} {\bibfnamefont {E.}~\bibnamefont
  {Demler}},\ }\bibfield  {title} {\bibinfo {title} {Measuring entanglement
  entropy of a generic many-body system with a quantum switch},\ }\href
  {https://doi.org/10.1103/PhysRevLett.109.020504} {\bibfield  {journal}
  {\bibinfo  {journal} {Phys. Rev. Lett.}\ }\textbf {\bibinfo {volume} {109}},\
  \bibinfo {pages} {020504} (\bibinfo {year} {2012})}\BibitemShut {NoStop}%
\bibitem [{\citenamefont {Calixto}\ and\ \citenamefont
  {Romera}(2015{\natexlab{a}})}]{silicene1}%
  \BibitemOpen
  \bibfield  {author} {\bibinfo {author} {\bibfnamefont {M.}~\bibnamefont
  {Calixto}}\ and\ \bibinfo {author} {\bibfnamefont {E.}~\bibnamefont
  {Romera}},\ }\bibfield  {title} {\bibinfo {title} {Identifying
  topological-band insulator transitions in silicene and other 2d gapped dirac
  materials by means of {Rényi-Wehrl} entropy},\ }\href
  {https://doi.org/10.1209/0295-5075/109/40003} {\bibfield  {journal} {\bibinfo
   {journal} {{EPL} (Europhysics Letters)}\ }\textbf {\bibinfo {volume}
  {109}},\ \bibinfo {pages} {40003} (\bibinfo {year}
  {2015}{\natexlab{a}})}\BibitemShut {NoStop}%
\bibitem [{\citenamefont {Calixto}\ and\ \citenamefont
  {Romera}(2015{\natexlab{b}})}]{silicene3}%
  \BibitemOpen
  \bibfield  {author} {\bibinfo {author} {\bibfnamefont {M.}~\bibnamefont
  {Calixto}}\ and\ \bibinfo {author} {\bibfnamefont {E.}~\bibnamefont
  {Romera}},\ }\bibfield  {title} {\bibinfo {title} {Inverse participation
  ratio and localization in topological insulator phase transitions},\ }\href
  {https://doi.org/10.1088/1742-5468/2015/06/p06029} {\bibfield  {journal}
  {\bibinfo  {journal} {Journal of Statistical Mechanics: Theory and
  Experiment}\ }\textbf {\bibinfo {volume} {2015}},\ \bibinfo {pages} {P06029}
  (\bibinfo {year} {2015}{\natexlab{b}})}\BibitemShut {NoStop}%
\bibitem [{\citenamefont {Castaños}\ \emph {et~al.}(2019)\citenamefont
  {Castaños}, \citenamefont {Romera},\ and\ \citenamefont {Calixto}}]{MRX}%
  \BibitemOpen
  \bibfield  {author} {\bibinfo {author} {\bibfnamefont {O.}~\bibnamefont
  {Castaños}}, \bibinfo {author} {\bibfnamefont {E.}~\bibnamefont {Romera}},\
  and\ \bibinfo {author} {\bibfnamefont {M.}~\bibnamefont {Calixto}},\
  }\bibfield  {title} {\bibinfo {title} {Information theoretic analysis of
  landau levels in monolayer phosphorene under magnetic and electric fields},\
  }\href {https://doi.org/10.1088/2053-1591/ab3fdc} {\bibfield  {journal}
  {\bibinfo  {journal} {Materials Research Express}\ }\textbf {\bibinfo
  {volume} {6}},\ \bibinfo {pages} {106316} (\bibinfo {year}
  {2019})}\BibitemShut {NoStop}%
\bibitem [{\citenamefont {Calixto}\ \emph {et~al.}(2021)\citenamefont
  {Calixto}, \citenamefont {Romera},\ and\ \citenamefont {Casta\~nos}}]{IJQC}%
  \BibitemOpen
  \bibfield  {author} {\bibinfo {author} {\bibfnamefont {M.}~\bibnamefont
  {Calixto}}, \bibinfo {author} {\bibfnamefont {E.}~\bibnamefont {Romera}},\
  and\ \bibinfo {author} {\bibfnamefont {O.}~\bibnamefont {Casta\~nos}},\
  }\bibfield  {title} {\bibinfo {title} {Analogies between the topological
  insulator phase of 2d dirac materials and the superradiant phase of
  atom-field systems},\ }\href {https://doi.org/10.1002/qua.26464} {\bibfield
  {journal} {\bibinfo  {journal} {International Journal of Quantum Chemistry}\
  }\textbf {\bibinfo {volume} {121}},\ \bibinfo {pages} {e26464} (\bibinfo
  {year} {2021})}\BibitemShut {NoStop}%
\bibitem [{\citenamefont {Calixto}\ \emph {et~al.}(2022)\citenamefont
  {Calixto}, \citenamefont {Cordero}, \citenamefont {Romera},\ and\
  \citenamefont {Castaños}}]{CALIXTO2022128057}%
  \BibitemOpen
  \bibfield  {author} {\bibinfo {author} {\bibfnamefont {M.}~\bibnamefont
  {Calixto}}, \bibinfo {author} {\bibfnamefont {N.~A.}\ \bibnamefont
  {Cordero}}, \bibinfo {author} {\bibfnamefont {E.}~\bibnamefont {Romera}},\
  and\ \bibinfo {author} {\bibfnamefont {O.}~\bibnamefont {Castaños}},\
  }\bibfield  {title} {\bibinfo {title} {Signatures of topological phase
  transitions in higher landau levels of {HgTe/CdTe} quantum wells from an
  information theory perspective},\ }\href
  {https://doi.org/https://doi.org/10.1016/j.physa.2022.128057} {\bibfield
  {journal} {\bibinfo  {journal} {Physica A: Statistical Mechanics and its
  Applications}\ }\textbf {\bibinfo {volume} {605}},\ \bibinfo {pages} {128057}
  (\bibinfo {year} {2022})}\BibitemShut {NoStop}%
\bibitem [{\citenamefont {Calixto}\ and\ \citenamefont
  {Castaños}(2024)}]{calixto2024IPRHgTe}%
  \BibitemOpen
  \bibfield  {author} {\bibinfo {author} {\bibfnamefont {M.}~\bibnamefont
  {Calixto}}\ and\ \bibinfo {author} {\bibfnamefont {O.}~\bibnamefont
  {Castaños}},\ }\href {https://arxiv.org/abs/2407.12469} {\bibinfo {title}
  {Inverse participation ratio and entanglement of edge states in {HgTe}
  quantum wells in a finite strip geometry}} (\bibinfo {year} {2024}),\ \Eprint
  {https://arxiv.org/abs/2407.12469} {arXiv:2407.12469 [cond-mat.mes-hall]}
  \BibitemShut {NoStop}%
\bibitem [{\citenamefont {Zanardi}\ and\ \citenamefont {{Paunkovi\ifmmode
  \acute{c}\else ć\fi{}}}(2006)}]{Zanardi}%
  \BibitemOpen
  \bibfield  {author} {\bibinfo {author} {\bibfnamefont {P.}~\bibnamefont
  {Zanardi}}\ and\ \bibinfo {author} {\bibfnamefont {N.}~\bibnamefont
  {{Paunkovi\ifmmode \acute{c}\else ć\fi{}}}},\ }\bibfield  {title} {\bibinfo
  {title} {Ground state overlap and quantum phase transitions},\ }\href
  {https://doi.org/10.1103/PhysRevE.74.031123} {\bibfield  {journal} {\bibinfo
  {journal} {Phys. Rev. E}\ }\textbf {\bibinfo {volume} {74}},\ \bibinfo
  {pages} {031123} (\bibinfo {year} {2006})}\BibitemShut {NoStop}%
\bibitem [{\citenamefont {Zanardi}\ \emph {et~al.}(2007)\citenamefont
  {Zanardi}, \citenamefont {Giorda},\ and\ \citenamefont
  {Cozzini}}]{PhysRevLett.99.100603}%
  \BibitemOpen
  \bibfield  {author} {\bibinfo {author} {\bibfnamefont {P.}~\bibnamefont
  {Zanardi}}, \bibinfo {author} {\bibfnamefont {P.}~\bibnamefont {Giorda}},\
  and\ \bibinfo {author} {\bibfnamefont {M.}~\bibnamefont {Cozzini}},\
  }\bibfield  {title} {\bibinfo {title} {Information-theoretic differential
  geometry of quantum phase transitions},\ }\href
  {https://doi.org/10.1103/PhysRevLett.99.100603} {\bibfield  {journal}
  {\bibinfo  {journal} {Phys. Rev. Lett.}\ }\textbf {\bibinfo {volume} {99}},\
  \bibinfo {pages} {100603} (\bibinfo {year} {2007})}\BibitemShut {NoStop}%
\bibitem [{\citenamefont {You}\ \emph {et~al.}(2007)\citenamefont {You},
  \citenamefont {Li},\ and\ \citenamefont {Gu}}]{GuPRE2007}%
  \BibitemOpen
  \bibfield  {author} {\bibinfo {author} {\bibfnamefont {W.-L.}\ \bibnamefont
  {You}}, \bibinfo {author} {\bibfnamefont {Y.-W.}\ \bibnamefont {Li}},\ and\
  \bibinfo {author} {\bibfnamefont {S.-J.}\ \bibnamefont {Gu}},\ }\bibfield
  {title} {\bibinfo {title} {Fidelity, dynamic structure factor, and
  susceptibility in critical phenomena},\ }\href
  {https://doi.org/10.1103/PhysRevE.76.022101} {\bibfield  {journal} {\bibinfo
  {journal} {Phys. Rev. E}\ }\textbf {\bibinfo {volume} {76}},\ \bibinfo
  {pages} {022101} (\bibinfo {year} {2007})}\BibitemShut {NoStop}%
\bibitem [{\citenamefont {Gu}(2010)}]{GuIJMPB2010}%
  \BibitemOpen
  \bibfield  {author} {\bibinfo {author} {\bibfnamefont {S.-J.}\ \bibnamefont
  {Gu}},\ }\bibfield  {title} {\bibinfo {title} {Fidelity approach to quantum
  phase transitions},\ }\href {https://doi.org/10.1142/S0217979210056335}
  {\bibfield  {journal} {\bibinfo  {journal} {International Journal of Modern
  Physics B}\ }\textbf {\bibinfo {volume} {24}},\ \bibinfo {pages} {4371}
  (\bibinfo {year} {2010})}\BibitemShut {NoStop}%
\bibitem [{\citenamefont {Casta{\~{n}}os}\ \emph {et~al.}(2012)\citenamefont
  {Casta{\~{n}}os}, \citenamefont {L{\'{o}}pez-Pe{\~{n}}a}, \citenamefont
  {Nahmad-Achar},\ and\ \citenamefont {Hirsch}}]{octaviofide}%
  \BibitemOpen
  \bibfield  {author} {\bibinfo {author} {\bibfnamefont {O.}~\bibnamefont
  {Casta{\~{n}}os}}, \bibinfo {author} {\bibfnamefont {R.}~\bibnamefont
  {L{\'{o}}pez-Pe{\~{n}}a}}, \bibinfo {author} {\bibfnamefont {E.}~\bibnamefont
  {Nahmad-Achar}},\ and\ \bibinfo {author} {\bibfnamefont {J.~G.}\ \bibnamefont
  {Hirsch}},\ }\bibfield  {title} {\bibinfo {title} {Quantum information
  approach to the description of quantum phase transitions},\ }\href
  {https://doi.org/10.1088/1742-6596/403/1/012003} {\bibfield  {journal}
  {\bibinfo  {journal} {Journal of Physics: Conference Series}\ }\textbf
  {\bibinfo {volume} {403}},\ \bibinfo {pages} {012003} (\bibinfo {year}
  {2012})}\BibitemShut {NoStop}%
\bibitem [{\citenamefont {Cordero}\ \emph {et~al.}(2021)\citenamefont
  {Cordero}, \citenamefont {Nahmad-Achar}, \citenamefont
  {L{\'{o}}pez-Pe{\~{n}}a},\ and\ \citenamefont
  {Casta{\~{n}}os}}]{octaviofide1}%
  \BibitemOpen
  \bibfield  {author} {\bibinfo {author} {\bibfnamefont {S.}~\bibnamefont
  {Cordero}}, \bibinfo {author} {\bibfnamefont {E.}~\bibnamefont
  {Nahmad-Achar}}, \bibinfo {author} {\bibfnamefont {R.}~\bibnamefont
  {L{\'{o}}pez-Pe{\~{n}}a}},\ and\ \bibinfo {author} {\bibfnamefont
  {O.}~\bibnamefont {Casta{\~{n}}os}},\ }\bibfield  {title} {\bibinfo {title}
  {Quantum phase diagrams of matter-field hamiltonians {I}: Fidelity, bures
  distance, and entanglement},\ }\href
  {https://doi.org/10.1088/1402-4896/abd653} {\bibfield  {journal} {\bibinfo
  {journal} {Physica Scripta}\ }\textbf {\bibinfo {volume} {96}},\ \bibinfo
  {pages} {035104} (\bibinfo {year} {2021})}\BibitemShut {NoStop}%
\bibitem [{\citenamefont {L{\'{o}}pez-Pe{\~{n}}a}\ \emph
  {et~al.}(2021)\citenamefont {L{\'{o}}pez-Pe{\~{n}}a}, \citenamefont
  {Cordero}, \citenamefont {Nahmad-Achar},\ and\ \citenamefont
  {Casta{\~{n}}os}}]{octaviofide2}%
  \BibitemOpen
  \bibfield  {author} {\bibinfo {author} {\bibfnamefont {R.}~\bibnamefont
  {L{\'{o}}pez-Pe{\~{n}}a}}, \bibinfo {author} {\bibfnamefont {S.}~\bibnamefont
  {Cordero}}, \bibinfo {author} {\bibfnamefont {E.}~\bibnamefont
  {Nahmad-Achar}},\ and\ \bibinfo {author} {\bibfnamefont {O.}~\bibnamefont
  {Casta{\~{n}}os}},\ }\bibfield  {title} {\bibinfo {title} {Quantum phase
  diagrams of matter-field hamiltonians {II}: Wigner function analysis},\
  }\href {https://doi.org/10.1088/1402-4896/abd654} {\bibfield  {journal}
  {\bibinfo  {journal} {Physica Scripta}\ }\textbf {\bibinfo {volume} {96}},\
  \bibinfo {pages} {035103} (\bibinfo {year} {2021})}\BibitemShut {NoStop}%
\bibitem [{\citenamefont {Predin}\ \emph {et~al.}(2016)\citenamefont {Predin},
  \citenamefont {Wenk},\ and\ \citenamefont {Schliemann}}]{PhysRevB.93.115106}%
  \BibitemOpen
  \bibfield  {author} {\bibinfo {author} {\bibfnamefont {S.}~\bibnamefont
  {Predin}}, \bibinfo {author} {\bibfnamefont {P.}~\bibnamefont {Wenk}},\ and\
  \bibinfo {author} {\bibfnamefont {J.}~\bibnamefont {Schliemann}},\ }\bibfield
   {title} {\bibinfo {title} {Trigonal warping in bilayer graphene: Energy
  versus entanglement spectrum},\ }\href
  {https://doi.org/10.1103/PhysRevB.93.115106} {\bibfield  {journal} {\bibinfo
  {journal} {Phys. Rev. B}\ }\textbf {\bibinfo {volume} {93}},\ \bibinfo
  {pages} {115106} (\bibinfo {year} {2016})}\BibitemShut {NoStop}%
\bibitem [{\citenamefont {Bittencourt}\ and\ \citenamefont
  {Bernardini}(2017)}]{PhysRevB.95.195145}%
  \BibitemOpen
  \bibfield  {author} {\bibinfo {author} {\bibfnamefont {V.~A. S.~V.}\
  \bibnamefont {Bittencourt}}\ and\ \bibinfo {author} {\bibfnamefont {A.~E.}\
  \bibnamefont {Bernardini}},\ }\bibfield  {title} {\bibinfo {title}
  {Lattice-layer entanglement in bernal-stacked bilayer graphene},\ }\href
  {https://doi.org/10.1103/PhysRevB.95.195145} {\bibfield  {journal} {\bibinfo
  {journal} {Phys. Rev. B}\ }\textbf {\bibinfo {volume} {95}},\ \bibinfo
  {pages} {195145} (\bibinfo {year} {2017})}\BibitemShut {NoStop}%
\bibitem [{\citenamefont {Catarina}\ \emph {et~al.}(2019)\citenamefont
  {Catarina}, \citenamefont {Amorim}, \citenamefont {Castro}, \citenamefont
  {Lopes},\ and\ \citenamefont {Peres}}]{Catarina2019}%
  \BibitemOpen
  \bibfield  {author} {\bibinfo {author} {\bibfnamefont {G.}~\bibnamefont
  {Catarina}}, \bibinfo {author} {\bibfnamefont {B.}~\bibnamefont {Amorim}},
  \bibinfo {author} {\bibfnamefont {E.~V.}\ \bibnamefont {Castro}}, \bibinfo
  {author} {\bibfnamefont {J.~M. V.~P.}\ \bibnamefont {Lopes}},\ and\ \bibinfo
  {author} {\bibfnamefont {N.}~\bibnamefont {Peres}},\ }\bibinfo {title}
  {Twisted bilayer graphene: Low-energy physics, electronic and optical
  properties},\ in\ \href
  {https://doi.org/https://doi.org/10.1002/9781119468455.ch44} {\emph {\bibinfo
  {booktitle} {Handbook of Graphene Set}}}\ (\bibinfo  {publisher} {John Wiley
  and Sons, Ltd},\ \bibinfo {year} {2019})\ Chap.~\bibinfo {chapter} {6}, pp.\
  \bibinfo {pages} {177--231}\BibitemShut {NoStop}%
\bibitem [{\citenamefont {Meneghini}(2020)}]{Beppe}%
  \BibitemOpen
  \bibfield  {author} {\bibinfo {author} {\bibfnamefont {G.}~\bibnamefont
  {Meneghini}},\ }\emph {\bibinfo {title} {Electronic properties of twisted
  bilayer graphene}},\ \href@noop {} {\bibinfo {type} {Master degree in
  physics}},\ \bibinfo  {school} {University of Padova}, \bibinfo {address}
  {Dipartimento di Fisica e Astronomia “Galileo Galilei”} (\bibinfo {year}
  {2020}),\ \bibinfo {note}
  {\url{https://hdl.handle.net/20.500.12608/22974}}\BibitemShut {NoStop}%
\bibitem [{\citenamefont {Schlienz}\ and\ \citenamefont
  {Mahler}(1995)}]{Mahler}%
  \BibitemOpen
  \bibfield  {author} {\bibinfo {author} {\bibfnamefont {J.}~\bibnamefont
  {Schlienz}}\ and\ \bibinfo {author} {\bibfnamefont {G.}~\bibnamefont
  {Mahler}},\ }\bibfield  {title} {\bibinfo {title} {Description of
  entanglement},\ }\href {https://doi.org/10.1103/PhysRevA.52.4396} {\bibfield
  {journal} {\bibinfo  {journal} {Phys. Rev. A}\ }\textbf {\bibinfo {volume}
  {52}},\ \bibinfo {pages} {4396} (\bibinfo {year} {1995})}\BibitemShut
  {NoStop}%
\bibitem [{\citenamefont {Mahler}\ and\ \citenamefont
  {Weberruss}(1995)}]{mahlerBook}%
  \BibitemOpen
  \bibfield  {author} {\bibinfo {author} {\bibfnamefont {G.}~\bibnamefont
  {Mahler}}\ and\ \bibinfo {author} {\bibfnamefont {V.}~\bibnamefont
  {Weberruss}},\ }\href {https://doi.org/10.1007/978-3-662-03176-6} {\emph
  {\bibinfo {title} {Quantum networks}}},\ \bibinfo {edition} {1st}\ ed.\
  (\bibinfo  {publisher} {Springer-Verlag Berlin Heidelberg},\ \bibinfo {year}
  {1995})\BibitemShut {NoStop}%
\bibitem [{\citenamefont {Peres}(1996)}]{PhysRevLett.77.1413}%
  \BibitemOpen
  \bibfield  {author} {\bibinfo {author} {\bibfnamefont {A.}~\bibnamefont
  {Peres}},\ }\bibfield  {title} {\bibinfo {title} {Separability criterion for
  density matrices},\ }\href {https://doi.org/10.1103/PhysRevLett.77.1413}
  {\bibfield  {journal} {\bibinfo  {journal} {Phys. Rev. Lett.}\ }\textbf
  {\bibinfo {volume} {77}},\ \bibinfo {pages} {1413} (\bibinfo {year}
  {1996})}\BibitemShut {NoStop}%
\bibitem [{\citenamefont {Horodecki}\ \emph {et~al.}(1996)\citenamefont
  {Horodecki}, \citenamefont {Horodecki},\ and\ \citenamefont
  {Horodecki}}]{HORODECKI19961}%
  \BibitemOpen
  \bibfield  {author} {\bibinfo {author} {\bibfnamefont {M.}~\bibnamefont
  {Horodecki}}, \bibinfo {author} {\bibfnamefont {P.}~\bibnamefont
  {Horodecki}},\ and\ \bibinfo {author} {\bibfnamefont {R.}~\bibnamefont
  {Horodecki}},\ }\bibfield  {title} {\bibinfo {title} {Separability of mixed
  states: necessary and sufficient conditions},\ }\href
  {https://doi.org/https://doi.org/10.1016/S0375-9601(96)00706-2} {\bibfield
  {journal} {\bibinfo  {journal} {Physics Letters A}\ }\textbf {\bibinfo
  {volume} {223}},\ \bibinfo {pages} {1} (\bibinfo {year} {1996})}\BibitemShut
  {NoStop}%
\bibitem [{\citenamefont {Horodecki}\ \emph {et~al.}(2009)\citenamefont
  {Horodecki}, \citenamefont {Horodecki}, \citenamefont {Horodecki},\ and\
  \citenamefont {Horodecki}}]{RevModPhys.81.865-Horodecki}%
  \BibitemOpen
  \bibfield  {author} {\bibinfo {author} {\bibfnamefont {R.}~\bibnamefont
  {Horodecki}}, \bibinfo {author} {\bibfnamefont {P.}~\bibnamefont
  {Horodecki}}, \bibinfo {author} {\bibfnamefont {M.}~\bibnamefont
  {Horodecki}},\ and\ \bibinfo {author} {\bibfnamefont {K.}~\bibnamefont
  {Horodecki}},\ }\bibfield  {title} {\bibinfo {title} {Quantum entanglement},\
  }\href {https://doi.org/10.1103/RevModPhys.81.865} {\bibfield  {journal}
  {\bibinfo  {journal} {Rev. Mod. Phys.}\ }\textbf {\bibinfo {volume} {81}},\
  \bibinfo {pages} {865–942} (\bibinfo {year} {2009})}\BibitemShut {NoStop}%
\bibitem [{\citenamefont {Carr}\ \emph {et~al.}(2017)\citenamefont {Carr},
  \citenamefont {Massatt}, \citenamefont {Fang}, \citenamefont {Cazeaux},
  \citenamefont {Luskin},\ and\ \citenamefont {Kaxiras}}]{PhysRevB.95.075420}%
  \BibitemOpen
  \bibfield  {author} {\bibinfo {author} {\bibfnamefont {S.}~\bibnamefont
  {Carr}}, \bibinfo {author} {\bibfnamefont {D.}~\bibnamefont {Massatt}},
  \bibinfo {author} {\bibfnamefont {S.}~\bibnamefont {Fang}}, \bibinfo {author}
  {\bibfnamefont {P.}~\bibnamefont {Cazeaux}}, \bibinfo {author} {\bibfnamefont
  {M.}~\bibnamefont {Luskin}},\ and\ \bibinfo {author} {\bibfnamefont
  {E.}~\bibnamefont {Kaxiras}},\ }\bibfield  {title} {\bibinfo {title}
  {Twistronics: Manipulating the electronic properties of two-dimensional
  layered structures through their twist angle},\ }\href
  {https://doi.org/10.1103/PhysRevB.95.075420} {\bibfield  {journal} {\bibinfo
  {journal} {Phys. Rev. B}\ }\textbf {\bibinfo {volume} {95}},\ \bibinfo
  {pages} {075420} (\bibinfo {year} {2017})}\BibitemShut {NoStop}%
\bibitem [{\citenamefont {Garreis}\ \emph {et~al.}(2024)\citenamefont
  {Garreis}, \citenamefont {Tong}, \citenamefont {Terle}, \citenamefont
  {Ruckriegel}, \citenamefont {Gerber}, \citenamefont {G{\"a}chter},
  \citenamefont {Watanabe}, \citenamefont {Taniguchi}, \citenamefont {Ihn},
  \citenamefont {Ensslin},\ and\ \citenamefont {Huang}}]{Garreis2024}%
  \BibitemOpen
  \bibfield  {author} {\bibinfo {author} {\bibfnamefont {R.}~\bibnamefont
  {Garreis}}, \bibinfo {author} {\bibfnamefont {C.}~\bibnamefont {Tong}},
  \bibinfo {author} {\bibfnamefont {J.}~\bibnamefont {Terle}}, \bibinfo
  {author} {\bibfnamefont {M.~J.}\ \bibnamefont {Ruckriegel}}, \bibinfo
  {author} {\bibfnamefont {J.~D.}\ \bibnamefont {Gerber}}, \bibinfo {author}
  {\bibfnamefont {L.~M.}\ \bibnamefont {G{\"a}chter}}, \bibinfo {author}
  {\bibfnamefont {K.}~\bibnamefont {Watanabe}}, \bibinfo {author}
  {\bibfnamefont {T.}~\bibnamefont {Taniguchi}}, \bibinfo {author}
  {\bibfnamefont {T.}~\bibnamefont {Ihn}}, \bibinfo {author} {\bibfnamefont
  {K.}~\bibnamefont {Ensslin}},\ and\ \bibinfo {author} {\bibfnamefont {W.~W.}\
  \bibnamefont {Huang}},\ }\bibfield  {title} {\bibinfo {title} {Long-lived
  valley states in bilayer graphene quantum dots},\ }\href
  {https://doi.org/10.1038/s41567-023-02334-7} {\bibfield  {journal} {\bibinfo
  {journal} {Nature Physics}\ }\textbf {\bibinfo {volume} {20}},\ \bibinfo
  {pages} {428} (\bibinfo {year} {2024})}\BibitemShut {NoStop}%
\bibitem [{\citenamefont {Park}\ \emph {et~al.}(2021)\citenamefont {Park},
  \citenamefont {Cao}, \citenamefont {Watanabe}, \citenamefont {Taniguchi},\
  and\ \citenamefont {Jarillo-Herrero}}]{Park2021}%
  \BibitemOpen
  \bibfield  {author} {\bibinfo {author} {\bibfnamefont {J.~M.}\ \bibnamefont
  {Park}}, \bibinfo {author} {\bibfnamefont {Y.}~\bibnamefont {Cao}}, \bibinfo
  {author} {\bibfnamefont {K.}~\bibnamefont {Watanabe}}, \bibinfo {author}
  {\bibfnamefont {T.}~\bibnamefont {Taniguchi}},\ and\ \bibinfo {author}
  {\bibfnamefont {P.}~\bibnamefont {Jarillo-Herrero}},\ }\bibfield  {title}
  {\bibinfo {title} {Tunable strongly coupled superconductivity in magic-angle
  twisted trilayer graphene},\ }\href
  {https://doi.org/10.1038/s41586-021-03192-0} {\bibfield  {journal} {\bibinfo
  {journal} {Nature}\ }\textbf {\bibinfo {volume} {590}},\ \bibinfo {pages}
  {249} (\bibinfo {year} {2021})}\BibitemShut {NoStop}%
\bibitem [{\citenamefont {Devakul}\ \emph {et~al.}(2023)\citenamefont
  {Devakul}, \citenamefont {Ledwith}, \citenamefont {Xia}, \citenamefont {Uri},
  \citenamefont {de~la Barrera}, \citenamefont {Jarillo-Herrero},\ and\
  \citenamefont {Fu}}]{HTG}%
  \BibitemOpen
  \bibfield  {author} {\bibinfo {author} {\bibfnamefont {T.}~\bibnamefont
  {Devakul}}, \bibinfo {author} {\bibfnamefont {P.~J.}\ \bibnamefont
  {Ledwith}}, \bibinfo {author} {\bibfnamefont {L.-Q.}\ \bibnamefont {Xia}},
  \bibinfo {author} {\bibfnamefont {A.}~\bibnamefont {Uri}}, \bibinfo {author}
  {\bibfnamefont {S.~C.}\ \bibnamefont {de~la Barrera}}, \bibinfo {author}
  {\bibfnamefont {P.}~\bibnamefont {Jarillo-Herrero}},\ and\ \bibinfo {author}
  {\bibfnamefont {L.}~\bibnamefont {Fu}},\ }\bibfield  {title} {\bibinfo
  {title} {Magic-angle helical trilayer graphene},\ }\href
  {https://doi.org/10.1126/sciadv.adi6063} {\bibfield  {journal} {\bibinfo
  {journal} {Science Advances}\ }\textbf {\bibinfo {volume} {9}},\ \bibinfo
  {pages} {1} (\bibinfo {year} {2023})}\BibitemShut {NoStop}%
\bibitem [{\citenamefont {Lu}\ \emph {et~al.}(2024)\citenamefont {Lu},
  \citenamefont {Han}, \citenamefont {Yao}, \citenamefont {Reddy},
  \citenamefont {Yang}, \citenamefont {Seo}, \citenamefont {Watanabe},
  \citenamefont {Taniguchi}, \citenamefont {Fu},\ and\ \citenamefont
  {Ju}}]{Lu2024}%
  \BibitemOpen
  \bibfield  {author} {\bibinfo {author} {\bibfnamefont {Z.}~\bibnamefont
  {Lu}}, \bibinfo {author} {\bibfnamefont {T.}~\bibnamefont {Han}}, \bibinfo
  {author} {\bibfnamefont {Y.}~\bibnamefont {Yao}}, \bibinfo {author}
  {\bibfnamefont {A.~P.}\ \bibnamefont {Reddy}}, \bibinfo {author}
  {\bibfnamefont {J.}~\bibnamefont {Yang}}, \bibinfo {author} {\bibfnamefont
  {J.}~\bibnamefont {Seo}}, \bibinfo {author} {\bibfnamefont {K.}~\bibnamefont
  {Watanabe}}, \bibinfo {author} {\bibfnamefont {T.}~\bibnamefont {Taniguchi}},
  \bibinfo {author} {\bibfnamefont {L.}~\bibnamefont {Fu}},\ and\ \bibinfo
  {author} {\bibfnamefont {L.}~\bibnamefont {Ju}},\ }\bibfield  {title}
  {\bibinfo {title} {Fractional quantum anomalous hall effect in multilayer
  graphene},\ }\href {https://doi.org/10.1038/s41586-023-07010-7} {\bibfield
  {journal} {\bibinfo  {journal} {Nature}\ }\textbf {\bibinfo {volume} {626}},\
  \bibinfo {pages} {759} (\bibinfo {year} {2024})}\BibitemShut {NoStop}%
\bibitem [{\citenamefont {Carr}\ \emph {et~al.}(2019)\citenamefont {Carr},
  \citenamefont {Fang}, \citenamefont {Zhu},\ and\ \citenamefont
  {Kaxiras}}]{PhysRevResearch.1.013001}%
  \BibitemOpen
  \bibfield  {author} {\bibinfo {author} {\bibfnamefont {S.}~\bibnamefont
  {Carr}}, \bibinfo {author} {\bibfnamefont {S.}~\bibnamefont {Fang}}, \bibinfo
  {author} {\bibfnamefont {Z.}~\bibnamefont {Zhu}},\ and\ \bibinfo {author}
  {\bibfnamefont {E.}~\bibnamefont {Kaxiras}},\ }\bibfield  {title} {\bibinfo
  {title} {Exact continuum model for low-energy electronic states of twisted
  bilayer graphene},\ }\href {https://doi.org/10.1103/PhysRevResearch.1.013001}
  {\bibfield  {journal} {\bibinfo  {journal} {Phys. Rev. Res.}\ }\textbf
  {\bibinfo {volume} {1}},\ \bibinfo {pages} {013001} (\bibinfo {year}
  {2019})}\BibitemShut {NoStop}%
\bibitem [{\citenamefont {Nguyen}\ \emph {et~al.}(2021)\citenamefont {Nguyen},
  \citenamefont {Paszko}, \citenamefont {Lamparski}, \citenamefont {Troeye},
  \citenamefont {Meunier},\ and\ \citenamefont {Charlier}}]{HungNguyen_2021}%
  \BibitemOpen
  \bibfield  {author} {\bibinfo {author} {\bibfnamefont {V.~H.}\ \bibnamefont
  {Nguyen}}, \bibinfo {author} {\bibfnamefont {D.}~\bibnamefont {Paszko}},
  \bibinfo {author} {\bibfnamefont {M.}~\bibnamefont {Lamparski}}, \bibinfo
  {author} {\bibfnamefont {B.~V.}\ \bibnamefont {Troeye}}, \bibinfo {author}
  {\bibfnamefont {V.}~\bibnamefont {Meunier}},\ and\ \bibinfo {author}
  {\bibfnamefont {J.-C.}\ \bibnamefont {Charlier}},\ }\bibfield  {title}
  {\bibinfo {title} {Electronic localization in small-angle twisted bilayer
  graphene},\ }\href {https://doi.org/10.1088/2053-1583/ac044f} {\bibfield
  {journal} {\bibinfo  {journal} {2D Materials}\ }\textbf {\bibinfo {volume}
  {8}},\ \bibinfo {pages} {035046} (\bibinfo {year} {2021})}\BibitemShut
  {NoStop}%
\bibitem [{\citenamefont {Xu}\ \emph {et~al.}(2019)\citenamefont {Xu},
  \citenamefont {Berdyugin}, \citenamefont {Kumaravadivel}, \citenamefont
  {Guinea}, \citenamefont {Krishna~Kumar}, \citenamefont {Bandurin},
  \citenamefont {Morozov}, \citenamefont {Kuang}, \citenamefont {Tsim},
  \citenamefont {Liu}, \citenamefont {Edgar}, \citenamefont {Grigorieva},
  \citenamefont {Fal'ko}, \citenamefont {Kim},\ and\ \citenamefont
  {Geim}}]{Xu2019}%
  \BibitemOpen
  \bibfield  {author} {\bibinfo {author} {\bibfnamefont {S.~G.}\ \bibnamefont
  {Xu}}, \bibinfo {author} {\bibfnamefont {A.~I.}\ \bibnamefont {Berdyugin}},
  \bibinfo {author} {\bibfnamefont {P.}~\bibnamefont {Kumaravadivel}}, \bibinfo
  {author} {\bibfnamefont {F.}~\bibnamefont {Guinea}}, \bibinfo {author}
  {\bibfnamefont {R.}~\bibnamefont {Krishna~Kumar}}, \bibinfo {author}
  {\bibfnamefont {D.~A.}\ \bibnamefont {Bandurin}}, \bibinfo {author}
  {\bibfnamefont {S.~V.}\ \bibnamefont {Morozov}}, \bibinfo {author}
  {\bibfnamefont {W.}~\bibnamefont {Kuang}}, \bibinfo {author} {\bibfnamefont
  {B.}~\bibnamefont {Tsim}}, \bibinfo {author} {\bibfnamefont {S.}~\bibnamefont
  {Liu}}, \bibinfo {author} {\bibfnamefont {J.~H.}\ \bibnamefont {Edgar}},
  \bibinfo {author} {\bibfnamefont {I.~V.}\ \bibnamefont {Grigorieva}},
  \bibinfo {author} {\bibfnamefont {V.~I.}\ \bibnamefont {Fal'ko}}, \bibinfo
  {author} {\bibfnamefont {M.}~\bibnamefont {Kim}},\ and\ \bibinfo {author}
  {\bibfnamefont {A.~K.}\ \bibnamefont {Geim}},\ }\bibfield  {title} {\bibinfo
  {title} {Giant oscillations in a triangular network of one-dimensional states
  in marginally twisted graphene},\ }\href
  {https://doi.org/10.1038/s41467-019-11971-7} {\bibfield  {journal} {\bibinfo
  {journal} {Nature Communications}\ }\textbf {\bibinfo {volume} {10}},\
  \bibinfo {pages} {4008} (\bibinfo {year} {2019})}\BibitemShut {NoStop}%
\bibitem [{\citenamefont {Yoo}\ \emph {et~al.}(2019)\citenamefont {Yoo},
  \citenamefont {Engelke}, \citenamefont {Carr}, \citenamefont {Fang},
  \citenamefont {Zhang}, \citenamefont {Cazeaux}, \citenamefont {Sung},
  \citenamefont {Hovden}, \citenamefont {Tsen}, \citenamefont {Taniguchi},
  \citenamefont {Watanabe}, \citenamefont {Yi}, \citenamefont {Kim},
  \citenamefont {Luskin}, \citenamefont {Tadmor}, \citenamefont {Kaxiras},\
  and\ \citenamefont {Kim}}]{Yoo2019}%
  \BibitemOpen
  \bibfield  {author} {\bibinfo {author} {\bibfnamefont {H.}~\bibnamefont
  {Yoo}}, \bibinfo {author} {\bibfnamefont {R.}~\bibnamefont {Engelke}},
  \bibinfo {author} {\bibfnamefont {S.}~\bibnamefont {Carr}}, \bibinfo {author}
  {\bibfnamefont {S.}~\bibnamefont {Fang}}, \bibinfo {author} {\bibfnamefont
  {K.}~\bibnamefont {Zhang}}, \bibinfo {author} {\bibfnamefont
  {P.}~\bibnamefont {Cazeaux}}, \bibinfo {author} {\bibfnamefont {S.~H.}\
  \bibnamefont {Sung}}, \bibinfo {author} {\bibfnamefont {R.}~\bibnamefont
  {Hovden}}, \bibinfo {author} {\bibfnamefont {A.~W.}\ \bibnamefont {Tsen}},
  \bibinfo {author} {\bibfnamefont {T.}~\bibnamefont {Taniguchi}}, \bibinfo
  {author} {\bibfnamefont {K.}~\bibnamefont {Watanabe}}, \bibinfo {author}
  {\bibfnamefont {G.-C.}\ \bibnamefont {Yi}}, \bibinfo {author} {\bibfnamefont
  {M.}~\bibnamefont {Kim}}, \bibinfo {author} {\bibfnamefont {M.}~\bibnamefont
  {Luskin}}, \bibinfo {author} {\bibfnamefont {E.~B.}\ \bibnamefont {Tadmor}},
  \bibinfo {author} {\bibfnamefont {E.}~\bibnamefont {Kaxiras}},\ and\ \bibinfo
  {author} {\bibfnamefont {P.}~\bibnamefont {Kim}},\ }\bibfield  {title}
  {\bibinfo {title} {Atomic and electronic reconstruction at the van der waals
  interface in twisted bilayer graphene},\ }\href
  {https://doi.org/10.1038/s41563-019-0346-z} {\bibfield  {journal} {\bibinfo
  {journal} {Nature Materials}\ }\textbf {\bibinfo {volume} {18}},\ \bibinfo
  {pages} {448} (\bibinfo {year} {2019})}\BibitemShut {NoStop}%
\end{thebibliography}%

\end{document}